\documentclass[%
twocolumn,
 amsmath,amssymb,
 aps,
]{revtex4-2}

\usepackage{graphicx}
\usepackage{dcolumn}
\usepackage{bm}

\usepackage[dvipsnames]{xcolor}
\usepackage{array}
\usepackage{multirow}
\usepackage{lipsum}
\def\GeV2{{\rm GeV}^2}

\begin{document}

\title{Neutrino Cross Sections: \\ Interface of shallow- and deep-inelastic scattering for collider neutrinos}%

\author{Yu Seon Jeong}
\email{yusjeong@cau.ac.kr}
\affiliation{%
Chung-Ang University, High Energy Physics Center, 
Dongjak-gu, Seoul 06974, Republic of Korea 
}%
\author{Mary Hall Reno}%
\email{mary-hall-reno@uiowa.edu}
\affiliation{University of Iowa, Department of Physics and Astronomy,
Iowa City, Iowa 52242, USA\vskip 0.1in}

\date{\today}

\begin{abstract}
Neutrino experiments in a Forward Physics Facility at the Large Hadron Collider can measure neutrino and antineutrino cross sections for energies up to a few TeV. For neutrino energies below 100 GeV, the inelastic cross section evaluations have contributions from weak structure functions at low momentum transfers and low hadronic final state invariant mass. To evaluate the size of these contributions to the neutrino cross section, we use a parametrization of the electron-proton structure function, adapted for neutrino scattering, augmented with a correction to account for the partial conservation of the axial vector current,  and normalized to structure functions evaluated at next-to-leading order in QCD, with target mass corrections and heavy quark corrections. We compare our results with other approaches to account for this kinematic region in neutrino cross section for energies between 10--1000 GeV on isoscalar nucleon and iron targets.
\end{abstract}

\maketitle


\section{Introduction} 
\label{sec:intro}
Cross sections for neutrino interactions with nucleons and nuclei are important ingredients to measurements of neutrino mixing angles in neutrino oscillation experiments and for searches for physics beyond the standard model (BSM) \cite{Lipari:2002at,Katori:2016yel,Farzan:2017xzy,Arguelles:2019xgp}. The Deep Underground Neutrino Experiment (DUNE) will intercept neutrinos and antineutrinos produced at Fermilab \cite{DUNE:2016evb} with energies up to more than 10 GeV to meet their physics objectives \cite{DUNE:2015lol}.
The cross sections for neutrino energies of a few GeV have contributions from quasi-elastic scattering (e.g., $\nu_\mu n\to \mu p$), resonant production (e.g., $\nu_\mu n\to \Delta^0\to \mu n\pi^+$) and inelastic scattering $\nu_\mu N\to \mu X$, where $N=(p+n)/2$ represents an isoscalar nucleon target and $X$ refers to all final state hadrons. 
The quasi-elastic and resonant regions of the neutrino cross section are the focus of much of the theory effort, however, inelastic scattering is also important for neutrino interactions at DUNE \cite{Ruso:2022qes}. 
More than 50\% of DUNE events will be inelastic scattering events with the hadronic final state invariant mass larger than the mass of the $\Delta$ resonance, $m_\Delta (\approx 1.4~{\rm GeV})$~\cite{SajjadAthar:2020nvy,Morfin:2022hxe}. 

Inelastic scattering can involve small momentum transfers $Q^2$. Quark-hadron duality \cite{Bloom:1970xb,Bloom:1971ye} connects structure functions in the resonance region with parton-model-based structure functions at higher $Q^2$ for inelastic scattering \cite{Melnitchouk:2005zr,Lalakulich:2006yn}. 
For $E_\nu=10$ GeV, the neutrino charged-current (CC) cross section can have a contribution of $\sim 20\%$ from $Q^2<1$ GeV$^2$ \cite{Feng:2022inv}, a $Q^2$ kinematic regime where a perturbative QCD parton distribution function decomposition of the weak structure functions cannot necessarily be performed.

The opportunity to measure neutrino and antineutrino cross sections and weak structure functions will arise in experiments in the forward region at the Large Hadron Collider (LHC). 
Recently, two neutrino experiments at the LHC, FASER$\nu$ \cite{FASER:2020gpr} and SND@LHC \cite{SHiP:2020sos,SNDLHC:2022ihg} have been installed in existing caverns along the beam-line direction, 480 m from the ATLAS interaction point, and are taking data. Both experiments reported their first observations of collider neutrino events with analysis of the data collected during 2022 \cite{FASER:2023zcr,SNDLHC:2023pun}. 
They will operate continuously during Run 3 of the LHC.
The next stage experiments for the high luminosity phase of the LHC  (HL-LHC) would run in the proposed Forward Physics Facility (FPF). The FPF would house larger neutrino detectors FASER$\nu$2, AdvSND and a liquid argon detector FLArE, and additional detectors designed for BSM searches \cite{Feng:2022inv,Anchordoqui:2021ghd}. 
In this configuration, a large number of neutrinos from the ATLAS interaction point emitted in the very forward region could be detected. 
With the increased luminosity and detector size at the FPF, the expected number of neutrino events is much larger than in FASER$\nu$ and SND@LHC. The estimated number of interactions are $\sim 10^6$ events for muon neutrinos and  thousands for tau neutrinos in FPF experiments \cite{Feng:2022inv}.

Neutrinos produced at the LHC in the forward region are distributed in energy up to a few TeV \cite{Feng:2022inv,Anchordoqui:2021ghd,Bai:2021ira,Bai:2022jcs,Kling:2021gos,Maciula:2022lzk,Bhattacharya:2023zei}. 
The TeV neutrino energy range has not been reached directly in neutrino beams produced by accelerators.
With a large number of neutrino events, the FPF can provide a unique opportunity to investigate neutrino interactions precisely at unexplored energy range for all three flavors \cite{Feng:2022inv,Candido:2023utz,Xie:2023suk}. 
Although a large fraction of neutrinos is distributed with energies above 100 GeV, also of interest are neutrinos with energies below 100 GeV. There will be thousands of neutrinos in the 10's of GeV energy range \cite{Feng:2022inv,Batell:2021aja} where cross sections are sensitive to low $Q^2$ structure functions.

In this work, we evaluate the charged-current neutrino deep-inelastic scattering (DIS) cross sections  
and investigate the size of the contribution from the transition region from resonance meson/baryon production to DIS, referred as the shallow inelastic scattering (SIS) region. 
We probe how the neutrino DIS cross sections depend on the hadronic final state invariant mass $W$ and on $Q^2$. We quantify the impact of ranges of these kinematic variables relevant to the SIS region for muon neutrino and tau neutrino as well as for antineutrino cross sections. In the next section, we outline approaches to low $Q^2$ extrapolations of electromagnetic structure functions and their translations to weak structure functions. In Section \ref{sec:pcac}, we introduce our approximation to the contributions from the partially conserved axial vector current.  Section \ref{sec:xc} shows our results for cross sections as a function of $W$ and $Q^2$, followed by a discussion in Section \ref{sec:conclusions}. We compare our structure function and cross section results with other approaches \cite{Bodek:2002ps,Bodek:2004pc,Bodek:2010km,Bodek:2021bde,Candido:2023utz}. For reference, the cross section formulas for neutrino and antineutrino charge-current cross sections are collected in Appendix A. Tables of cross section results are included in Appendix B.

\section{structure functions at Low $Q^2$}
\label{sec:SF}
In evaluating the cross section of neutrino charged-current (CC) DIS interaction, essential components are structure functions $F_{i,{\rm CC}} (x, Q^2)$ for $i=1-5$, as shown in detail in Appendix A. The structure functions are described by the parton momentum fraction $x$ (or Bjorken variable) and parton distribution functions (PDFs) $q(x,Q^2)$ and $\bar{q}(x,Q^2)$ for quarks and antiquarks labeled by their flavors, written schematically as
\begin{eqnarray}
\label{eq:f2cc}
F_{2,{\rm CC}} (x, Q^2) &=& \sum_{q,q'} 2 x \bigl(q(x,Q^2) +\bar{q}'(x,Q^2)\bigr)\, ,\\
\label{eq:f3cc}
F_{3,{\rm CC}} (x, Q^2) &=& \sum_{q,q'} 2 \bigl(q(x,Q^2) -\bar{q}'(x,Q^2)\bigr)\, ,
\end{eqnarray}
at leading order in perturbative QCD for $F_{2,{\rm CC}}$ and $F_{3,{\rm CC}}$. 
The sum is over the relevant quark flavors for neutrino and antineutrino scattering.
For more details for the distinction between neutrino and antineutrino structure functions, the reader is referred to Appendix A. By comparison, the electromagnetic structure function $F_{2,{\rm EM}}(x,Q^2)$ in the same approximation is
\begin{equation}
\label{eq:f2em}
    F_{2,{\rm EM}}= \sum_q e_q^2 x\bigl(q(x,Q^2) +\bar{q}(x,Q^2)\bigr)\,.
\end{equation}

PDFs are extracted from analyses of lepton-hadron and hadron-hadron data for ranges of the variables $x$ and $Q^2$ \cite{ParticleDataGroup:2022pth}. 
For the nCTEQ15 PDFs \cite{Kovarik:2015cma} used in this work, the ranges of applicability of the PDFs are $5 \times 10^{-6} < x < 1$ and $1.3~{\rm GeV} < Q < 10^4~{\rm GeV}$.
Since perturbative QCD is not applicable for $Q^2 \lesssim 1~{\rm GeV^2}$, PDF-based structure functions are not reliable in this low $Q^2$ region. 
The neutrino cross sections are sensitive to such low $Q^2$ at low incident neutrino energy, $E_\nu \lesssim {\cal O}(10)$~GeV, whereas the contributions from $Q^2 \lesssim 1~{\rm GeV^2}$ are negligible for neutrino energies above several hundred GeV. This paper shows results for neutrino and antineutrino cross sections from low $Q^2$ extrapolations of the CC structure functions for incident energies between $10-1000$ GeV.

In a discussion of the structure functions, it is useful to relate $x$ and $Q^2$ to the hadronic final state invariant mass $W$ defined by
\begin{equation}
\label{eq:w2}
    W^2=Q^2\Biggl( \frac{1}{x} -1\Biggr) + m_N^2\, ,
\end{equation}
where $m_N$ is the mass of the nucleon.
The DIS region is typically defined in terms of the $W$ and $Q^2$ as $W \gtrsim $ 2 GeV and $Q^2 \gtrsim 1 -4 $ GeV$^2$. 
One can consider the SIS region as $m_N + m_\pi < W \lesssim 2$ GeV for all $Q^2$ \cite{SajjadAthar:2020nvy}. 
However, the boundary between the SIS and DIS regions is not clearly defined.
Inelastic scattering that includes $W>W_{\rm min}=1.4$ GeV includes resonances above the $\Delta$ resonance. In the GENIE Monte Carlo generator, the value of $W_{\rm min}$ is set to 1.7~GeV~\cite{GENIE:2021npt,GENIE:2021zuu}.

In the context of the parton model for the lower range of the applicable $Q^2$, target mass corrections (TMC) modify the simplified forms of eqns. (\ref{eq:f2cc}) and (\ref{eq:f3cc}) \cite{Georgi:1976ve,Barbieri:1976rd,Kretzer:2003iu,Schienbein:2007gr,Ruiz:2023ozv}. Among the modifications is a replacement of the Bjorken-$x$ variable in the PDFs with the Nachtmann variable $\xi$ defined as
\begin{equation}
    \xi = \frac{2x}{1+\sqrt{1+4x^2 m_N^2/Q^2} }\,.
\end{equation}
The Nachtmann variable corresponds to the light-front momentum fraction of the struck parton in the massless parton limit, given the  massive nucleon target. 
When $m_N\to 0$, $\xi\to x$.
Target mass corrections to the parton model structure functions are largest for low $Q^2$ and for large $x$. The recent work on the TMC for nuclear targets \cite{Ruiz:2023ozv} shows that these corrections are effectively independent of target mass number. We use here the TMC for structure functions as outlined in refs. \cite{Kretzer:2003iu,Schienbein:2007gr}.

In the non-perturbative regime, the structure functions can be phenomenologically modeled by fitting to lepton-hadron scattering data.
There exist several models for such low $Q^2$ structure functions, e.g., refs.~\cite{Bodek:2004pc,Bodek:2021bde,Candido:2023utz,Capella:1994cr,Kaidalov:1998pn}. 
A commonly used model is the Bodek-Yang model \cite{Bodek:2004pc}, which provides effective PDFs at low $Q^2$ that can be used to construct electromagnetic and weak structure functions according to, e.g., eqns. (\ref{eq:f2cc}-\ref{eq:f2em}). The Bodek-Yang approach uses GRV98 LO PDFs \cite{Gluck:1998xa} evaluated with a modified Nachtmann variable and frozen at $Q_0^2=0.8$ GeV$^2$, then multiplied by $Q^2$ dependent $K$-factors. The parameters in the Bodek-Yang model are obtained by fitting
inelastic charged lepton-nucleon scattering data \cite{Whitlow:1991uw,BCDMS:1989ggw,NewMuon:1996fwh}.
The Bodek-Yang model is implemented in Monte Carlo event generators such as GENIE~\cite{GENIE:2021zuu}, NEUT~\cite{Hayato:2009zz} and NuWro~\cite{Juszczak:2005zs}.

\begin{figure} 
\centering
\includegraphics[width=.45\textwidth]{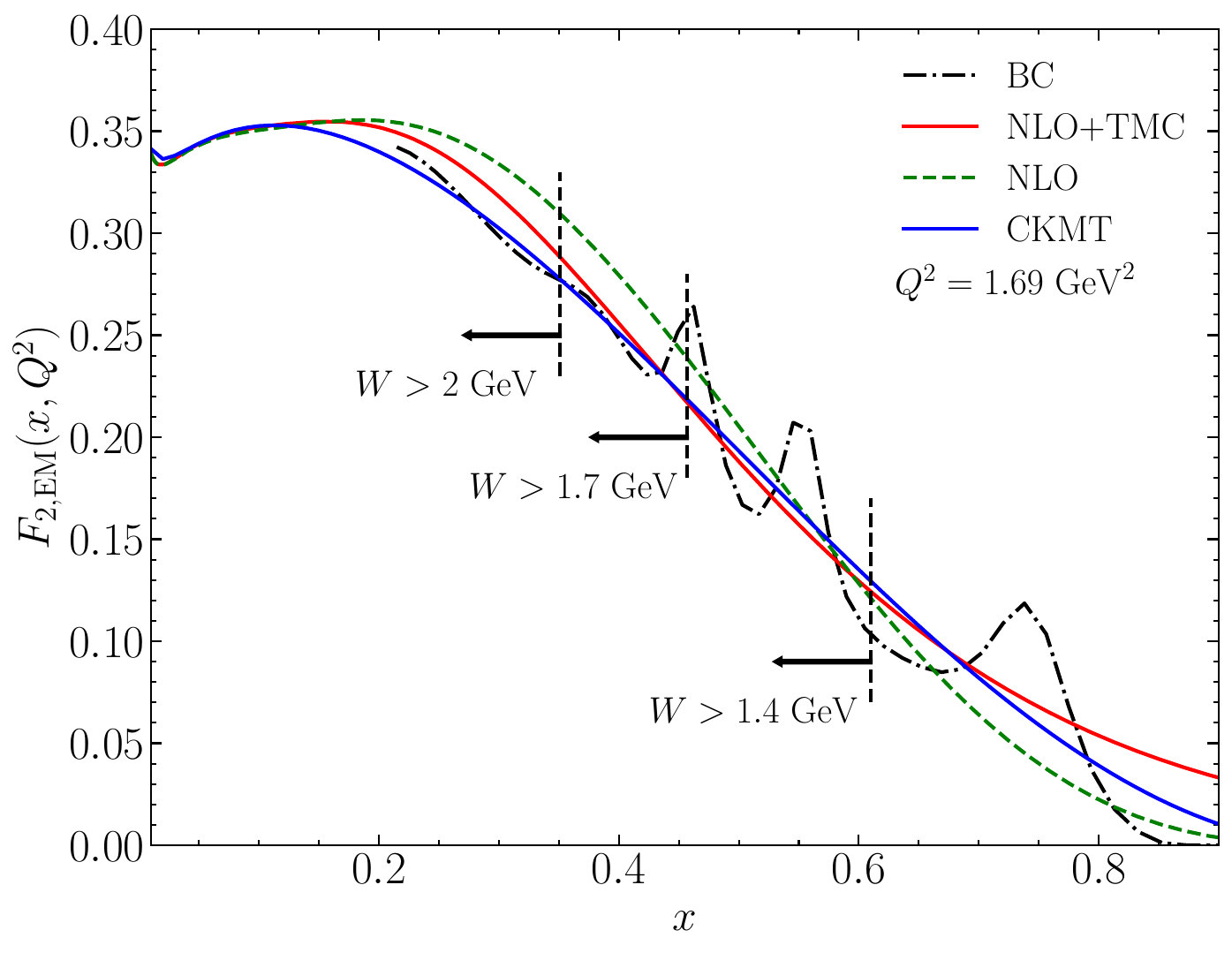}
\includegraphics[width=.45\textwidth]{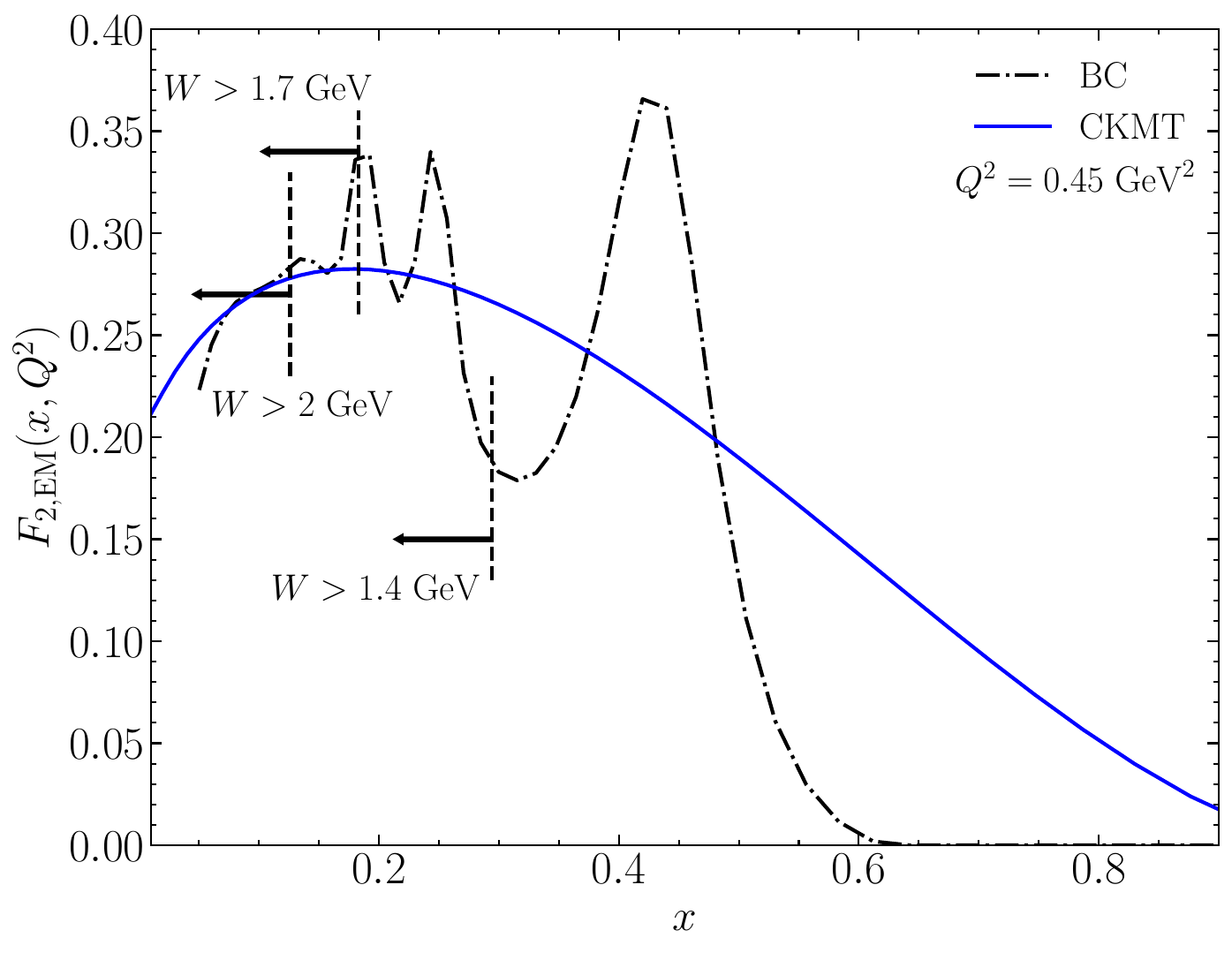}
\caption{\label{fig:f2compare} 
The structure functions $F_2 (x, Q^2)$ for electron-nucleon electromagnetic scattering for $Q^2$ = 1.69 GeV$^2$ (upper) and $0.45$ GeV$^2$ (lower). The CKMT \cite{Capella:1994cr}, NLO and NLO+TMC structure functions in the upper panel are compared with a parameterization of the electromagnetic scattering data by Bosted and Christy (BC) \cite{Christy:2007ve}. The lower panel shows only the BC and CKMT parameterizations.
}
\end{figure}

Another approach and the one that we use here starts with the parameterization of the structure function $F_{2,{\rm EM}}$ provided by Capella et al. (CKMT) in ref.~\cite{Capella:1994cr}. The CKMT parameterization \cite{Capella:1994cr,Kaidalov:1998pn}, obtained by fitting to electromagnetic structure function data, has the form
\begin{eqnarray}
\nonumber
F_{2,{\rm EM}}^{\rm CKMT}(x,Q^2)&=& 
F_2^{sea}(x,Q^2)+F_2^{val}(x,Q^2) \\ \nonumber
&=&Ax^{-\Delta(Q^2)}(1-x)^{n(Q^2)+4}\\ \nonumber&\times &\Biggl(
\frac{Q^2}{Q^2+a}\Biggr)^{1+\Delta(Q^2)}\\ \nonumber
&+& Bx^{1-\alpha_R}(1-x)^{n(Q^2)}\Biggl( \frac{Q^2}{Q^2+b}\Biggr)^{\alpha_R}\\ 
&\times & \Bigl( 1+ f(1-x)\Bigr) \ .
\label{eq:ckmt}
\end{eqnarray}
The functions $n(Q^2)$ and $\Delta (Q^2)$ are 
\begin{eqnarray}
n(Q^2)&=& \frac{3}{2}\Biggl( 1+\frac{Q^2}{Q^2+c}\Biggr)\ ,
\\
\Delta(Q^2)&=& \Delta_0 \Biggl( 1+\frac{2 Q^2}{Q^2+d}\Biggr)\ .
\end{eqnarray}
The constants entering into the parameterization of the electromagnetic structure function are in the first two rows of Table 1.

The functional form of eq. (\ref{eq:ckmt}) accounts for terms based on Pomeron and Reggeon contributions. At low $Q^2$, the distinction between sea and valence contributions is not evident, but at large enough $Q^2$, a parton model interpretation has the first term to be interpreted as the sea quark (and antiquark) contribution and the second term as the valence contribution.

In electromagnetic scattering, in the limit of $Q^2\to 0$ with $W^2$ fixed, the real photon-proton cross section in the CKMT form is
\begin{eqnarray}
    \sigma_{\gamma p}^{\rm tot}(W^2) &=& \frac{4\pi^2\alpha_{\rm EM}}{Q^2} F_{2,{\rm EM}}^{\rm CKMT}(x,Q^2)\lvert_{Q^2\to 0}\\
    \nonumber
    &\simeq & 4 \pi^2\alpha_{\rm EM}\Biggl[
    \frac{A}{a}\Biggl( \frac{W^2}{a}\Biggr)^{\Delta_0}\\
    \nonumber &+& \frac{B(1+f)}{b}
\Biggl(\frac{W^2}{b}\Biggr)^{\alpha_R-1}\Biggr]\,,
\end{eqnarray}
where $W^2\simeq Q^2/x$ for $W^2\gg m_N^2$. 
This means that $F_{2,EM}\sim Q^2$ for low $Q^2$.  

The upper panel of Fig. \ref{fig:f2compare} shows $F_{2,{\rm EM}}(x,Q^2)$ for $Q^2=1.69\ \GeV2$. The dot-dashed curve shows the parameterization  of Jefferson Laboratory electromagnetic structure function data by Bosted and Christy (BC) \cite{Christy:2007ve}. The dashed line show the next-to-leading order (NLO) QCD evaluation of the electromagnetic structure function using the nCTEQ15 proton PDF. The curve labeled NLO+TMC also includes target mass  \cite{Georgi:1976ve,Barbieri:1976rd,DeRujula:1976baf}.
The curve labeled CKMT shows the result using the parameterization of eq. (\ref{eq:ckmt}). In the lower panel of Fig. \ref{fig:f2compare}, for $Q^2=0.45$ GeV$^2$ where the PDF approach is not applicable, only the BC and CKMT parameterizations of $F_{2,{\rm EM}}$ are shown. 

To the left of vertical dashed lines in Fig. \ref{fig:f2compare} are the regions of $x$ such that $W$ is larger than the three values $W_{\rm min}=2,\ 1.7$ and 1.4 GeV given $Q^2=1.69\ \GeV2$ or $Q^2=0.45\ \GeV2$.
In the upper panel, one can see that the NLO perturbative result overshoots the data for $x\simeq 0.2-0.5$, while the NLO+TMC does somewhat better in the same range. The CKMT parametrization agrees well with the BC parameterization for $W>2$ GeV, and roughly averages the resonance region for smaller values of $W$. The lower panel shows that the CKMT structure function also roughly averages the structure function in the resonance region at lower $Q^2$, here shown for $Q^2=0.45$ GeV$^2$.

For CC scattering in the Bodek-Yang approach, the effective PDFs are assembled according to eqs. (\ref{eq:f2cc}) and (\ref{eq:f3cc}).
For $F_{1,{\rm CC}}(x,Q^2)$ we use the relation 
\begin{equation}
\label{eq:R}
R(x,Q^2)=\frac{F_{2,{\rm CC}}(x,Q^2)}{2xF_{1,{\rm CC}}(x,Q^2}\Biggl( 1+\frac{4m_N^2x^2}{Q^2}\Biggr) -1 
\end{equation}
to write $F_{1,{\rm CC}}$ in terms of 
$F_{2,{\rm CC}}$ and $R$, where a parameterization of $R$ is used. 

For CC scattering in the CKMT approach, the same functional form as $F_{2,{\rm EM}}$ is applied to $F_{2,{\rm CC}}$, with modifications to $A$, $B$ and $f$ to reflect the differences between eqs. (\ref{eq:f2cc})
and (\ref{eq:f2em}) for the sea, overall valence and relative importance of $u$ and $d$ valence contributions, respectively \cite{Reno:2006hj,Jeong:2010nt}.
The functional form for $xF_{3,{\rm CC}}$ is the same as for $F_{2,{\rm CC}}$, with the substitutions for $A$ and $B$ as indicated in Table 1.
For $F_{1,{\rm CC}}(x,Q^2)$, we use the relation in eq. (\ref{eq:R}) with 
the parameterization of Whitlow et al. \cite{Whitlow:1990gk} from electromagnetic scattering, which applies for $Q^2>Q_m^2=0.3\ \GeV2$.
Below $Q^2=Q_m^2$, we take $R(x,Q^2)=R(x,Q_m^2)\cdot Q^2/Q_m^2$. 

For $\nu_\tau$ and $\bar\nu_\tau$ charged current scattering, in principle two additional structure functions are required: $F_{4,{\rm CC}}$ and $F_{5,{\rm CC}}$. At leading order in QCD for massless partons, 
\begin{eqnarray}
\label{eq:f4}
    F_{4,{\rm CC}}(x,Q^2) &=& 0\,,\\
    \label{eq:f5}
    F_{5,{\rm CC}}(x,Q^2) &=& F_{2,{\rm CC}}(x,Q^2)/(2 x)\,.
\end{eqnarray}
These are the so-called Albright-Jarlskog relations \cite{Albright:1974ts}. 
We take into account the NLO QCD, target mass and heavy quark mass corrections to $F_{4,CC}$ and $F_{5,CC}$.  For the low-$Q^2$ extrapolations, following the Albright-Jarlskog relations, we set $F_{4,CC}$ =0 and extrapolate $F_5 (Q^2<Q_0^2)$ with the functional form of the $F_2(Q^2<Q_0^2)$ extrapolation discussed below.

\begin{figure} [t]
\centering
\includegraphics[width=.45\textwidth]{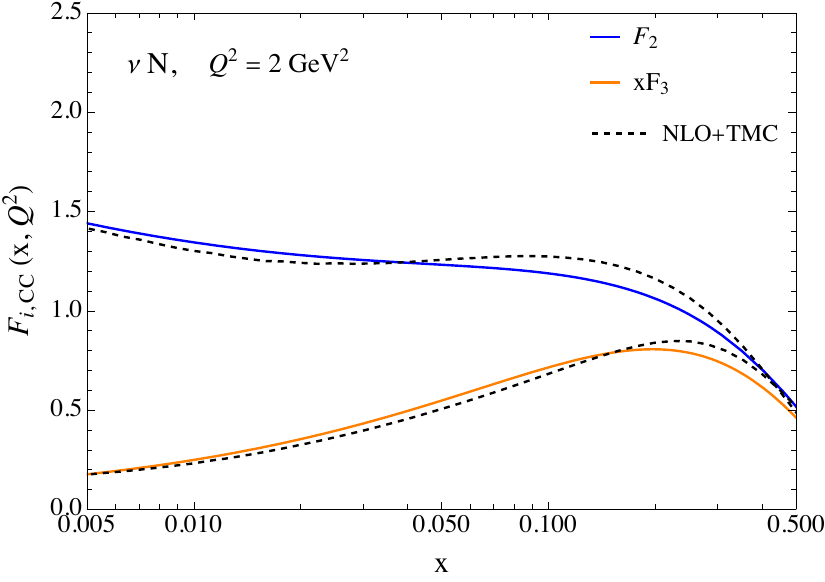}
\includegraphics[width=.45\textwidth]{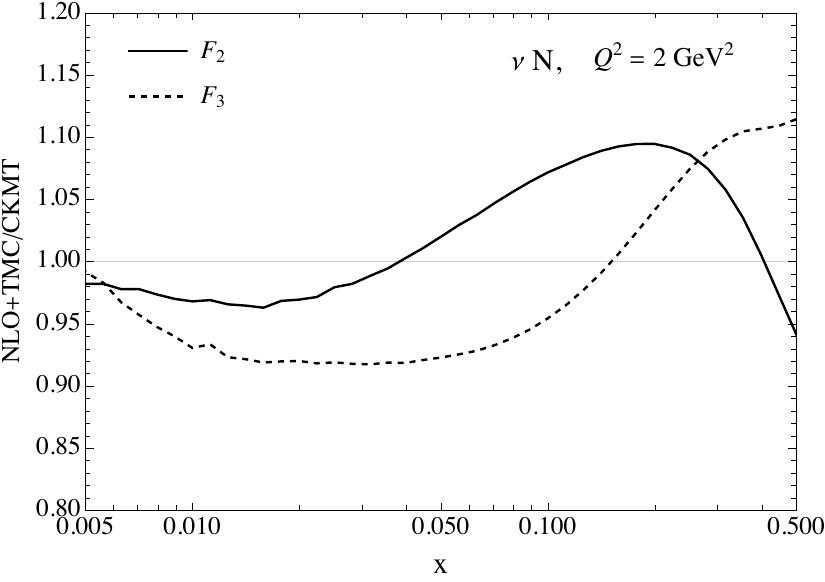}
\caption{\label{Fig:FiRatios} 
Upper: Comparison of the NLO+TMC (dashed) and CKMT (solid) evaluations of $F_{2,{\rm CC}}$ and $xF_{3,{\rm CC}}$. Lower:
The ratios of structure functions evaluated using NLO+TMC to CKMT for $Q^2=2\ \GeV2$.
}
\end{figure}

In our evaluations of the neutrino and antineutrino cross section, we use NLO+TMC evaluations of the structure functions that include 
heavy quark mass corrections \cite{Gottschalk:1980rv,Kretzer:2002fr,Kretzer:2003iu,Jeong:2010za} and
nCTEQ15 PDFs for $Q^2\geq Q_0^2$. For $Q^2<Q_0^2$, the CKMT neutrino structure functions are used. We will take $Q_0^2=2,\ 4$ GeV$^2$.
Comparisons of NLO+TMC and CKMT evaluations of $F_{2,{\rm CC}}$ and $xF_{3,{\rm CC}}$ at $Q^2=2$ GeV$^2$ are shown in the upper panel of Fig.~\ref{Fig:FiRatios}.  
Numerically, the CKMT structure function $F_{2,{\rm CC}}$ 
at $Q^2=2\ \GeV2$ is within $+9\%$ to $ -6\%$ of the NLO+TMC corrected structure function evaluated using the nCTEQ15 PDFs for $x<0.5$. 
The discrepancy is similar for $xF_{3,{\rm CC}}$, within $+11\%$ to $  - 8\%$. 

For reference, in the cross section evaluations below, the $xF_{3,{\rm CC}}$ contribution to the $\nu_\mu N$ CC cross section is $\sim 10\%$ for $E_\nu=10$ GeV, reducing to $\sim 0.4 \%$ for $E_\nu=10^2$ GeV. 
For $\bar\nu_\mu N$ CC scattering, the impact of $xF_{3,{\rm CC}}$ is larger because of the minus sign in the differential cross section (see Appendix A). 
Regarding the range of $x$ kinematically allowed for $Q^2>0.1$ GeV$^2$, for $E_\nu=10$ GeV, $x\gtrsim 6\times 10^{-3}$  in $\nu_\mu N$ CC scattering and $x\gtrsim 0.2$  in $\nu_\tau N$ CC scattering \cite{Reno:2021hrj}.
For $E_\nu\lesssim 1$ TeV, relevant to the neutrinos that can be detected at the FPF, the neutrino CC cross sections mostly come from $x\gtrsim 0.01$ \cite{Bai:2021ira}. 
For example, when $E_\nu = 1$ TeV, the contribution of $x > 0.01$ is about 95\% of the $\bar{\nu}_{\mu} N$ cross section, and it is higher for the $\nu_{\mu}$, $\nu_{\tau}$ and $\bar{\nu}_{\tau}$ cross sections.

To avoid discontinuities in the structure functions at $Q_0^2$, we normalize the CKMT structure functions according to
\begin{equation}
\label{eq:norm}
    F_{i,{\rm CC}}(x,Q^2)= F_{i,{\rm CC}}^{\rm CKMT}(x,Q^2)
    \frac{F_{i,{\rm CC}}^{\rm NLO+TMC}(x,Q_0^2)}{F_{i,{\rm CC}}^{\rm CKMT}(x,Q_0^2)}\,,
\end{equation}
and we label the CKMT structure functions normalized with NLO+TMC at $Q_0^2$ by CKMT-NT. As noted above, we use the Albright-Jarlskog relation to write $F_{5,{\rm CC}}^{\rm CKMT}$ in terms of $F_{2,{\rm CC}}^{\rm CKMT}/(2x)$ to evaluate $F_{5,{\rm CC}}$ for $Q^2<Q_0^2$. The normalization factors for $F_{2,{\rm CC}}$ and $xF_{3,{\rm CC}}$
are shown in the lower panel of Fig.~\ref{Fig:FiRatios} for $Q_0^2=2$ GeV$^2$.
Our results are not very sensitive to a choice of $Q_0^2$ between $2\ \GeV2$ and $4\ \GeV2$ because the CKMT parameterization follows the $Q^2$ evolution of $F_{2,{\rm CC}}$ reasonably well. 
As we show below, the results of $F_{2,{\rm CC}}$ evaluated at NLO+TMC matched to the low-$Q$ extrapolation for $Q_0^2=2$ GeV$^2$ and $Q_0^2=4$ GeV$^2$ according to eq. (\ref{eq:norm}) agree within $-6\%$ to $+2\%$ for $Q^2 \gtrsim 1$ GeV$^2$
(see Fig. \ref{Fig:F2pcac}). For $Q^2=0.1$ GeV$^2$, $F_{2,{\rm CC}}$ differs by up to $\sim 10\%$, however, because $F_{2,{\rm CC}}^{\rm CKMT}\to 0$ as $Q^2\to 0$, this difference hardly impacts the cross section evaluation.

Recently in ref. \cite{Candido:2023utz}, low $Q^2$ CC structure functions (NNSF$\nu$) were presented, modeled by a machine-learning parametrization of experimental data for neutrino  and antineutrino scattering on Ne, Fe, CaCO$_3$ and Pb. Structure functions for $F_{2,{\rm CC}},\ xF_{3,{\rm CC}}$ and $F_{L,{\rm CC}}$ are provided in LHAPDF format, separately with low-$Q$ and high-$Q$ grids \cite{Buckley:2014ana}.
The low-$Q$ grid is intended for $Q^2 \lesssim 500~{\rm GeV}^2$ and $x \geq 10^{-3}$, while the high-$Q$ grid extends to $Q^2$ outside this region.
The low-$Q$ grid is thus suitable for the $x$ range for cross sections for $E_\nu \lesssim 1$~TeV.
Contributions from $Q^2 > 500~{\rm GeV}^2$  are less than 8\% for both $\nu_\mu N$ and $\nu_\tau N$ cross sections at $E_\nu = 1$~TeV, and less than $\sim 1$\% for their antineutrino cross sections.
Therefore, we use only the low-$Q$ grids for the NNSF structure functions in our evaluations, and we label the results for deuterium structure functions by NNSF$\nu$(D). 
For reference, the cross sections evaluated using the low-$Q$ grid and the high-$Q$ grid agree within 1\% at 10 TeV, and the discrepancy becomes 2-3\% at 1 TeV.

\begin{figure*} [t]
\centering
\includegraphics[width=.49\textwidth]{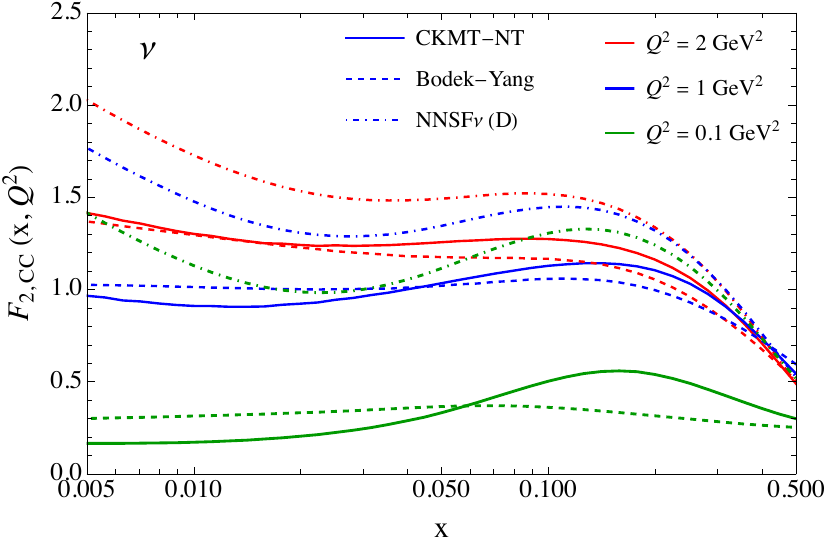}
\includegraphics[width=.49\textwidth]{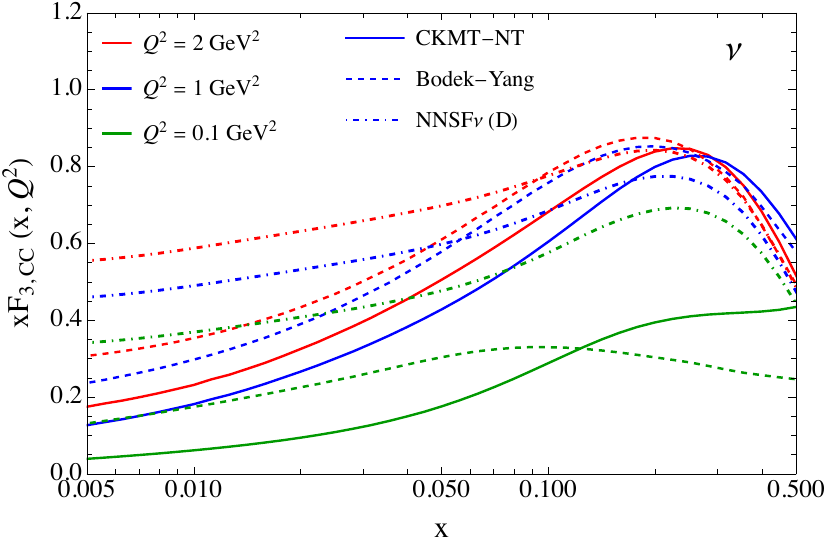}
\caption{
The structure functions $F_{2,{\rm CC}} (x, Q^2)$ (left) and $xF_{3,{\rm CC}} (x,Q^2)$ (right) for neutrino-nucleon charged current scattering for $Q^2$ = 0.1, 1 and 2 $\rm GeV^2$. The parameterizations of the neutrino CC structure functions from CKMT-NT \cite{Capella:1994cr, Reno:2006hj}, Bodek-Yang \cite{Bodek:2004pc} and NNSF$\nu$(D)
\cite{Candido:2023utz} are presented for comparison. The CKMT-NT structure functions are normalized at $Q_0^2$ = 2 GeV$^2$ in the curves shown here.}
\label{Fig:F2} 
\end{figure*}

\begin{table}
\begin{tabular}{cccccc}
\hline
\hline
$\Delta_0$ & $\alpha_R$ & $a\ [\GeV2 ]$ & $b\ [\GeV2 ]$ & $c\ [\GeV2 ]$ & $d\ [\GeV2 ]$ \\
0.07684  & 0.4150 & 0.2631 & 0.6452 & 3.5489 & 1.1170\\
\hline
\hline
 & Process & $A$ & $B$ & $f$ &\\
&EM $F_2$ & 0.1502  & 1.2064 & 0.15 &  \\
& $\nu N$ $F_2$ & 0.5967 & 2.7145 & 0.5962 & \\
& $\nu N$ $x F_3$ & 9.3955 $\times 10^{-3}$ &  2.4677 & 0.5962 & \\
& $\bar{\nu} N$ $x F_3$ & 9.3955 $\times 10^{-3}$ &  $-$2.4677 & 0.5962 & \\
\hline\hline
\end{tabular}
\caption{Parameter values in Ref. \cite{Kaidalov:1998pn} for CKMT parameterization of
the electromagnetic structure function $F_2$. The quantities $B$ and $f$ are determined
from the valence conditions at $Q^2=2\ \GeV2$.}
\end{table}

The two panels of Fig.~\ref{Fig:F2} show a comparison of the two structure functions 
$F_{2,{\rm CC}}$ and $xF_{3,{\rm CC}}$ for low $Q^2$ from the normalized CKMT-NT for neutrino CC interactions, the Bodek-Yang model,  and the  NNSF$\nu$(D) structure functions, each for several choices of $Q^2$: 0.1, 1 and 2 ${\rm GeV^2}$. 
One can see that the results from Bodek-Yang and CKMT-NT are comparable except for $Q^2 = 0.1~{\rm GeV^2}$.
The NNSF$\nu$(D) structure functions are consistently larger than Bodek-Yang and CKMT-NT, especially for $Q^2=0.1\ \GeV2$. Both the Bodek-Yang and CKMT-NT parameterizations have $F_{2,{\rm CC}}\to 0$ as $Q^2\to 0$ as in electromagnetic interactions, while the NNSF$\nu$ approach does not a priori require this. In fact, modifications of $F_{2,{\rm CC}}$ due to partial conservation of the axial vector current (PCAC) change the $Q^2\to 0$ limit, as we discuss below.

\section{Partially conserved axial vector current}
\label{sec:pcac}

There are recent investigations into modifications based on PCAC
\cite{Kulagin:2007ju,Bodek:2010km,Bodek:2021bde}, for which the weak structure function $F_{2,{\rm CC}}$ does not have the same $Q^2\to 0$ limit as the corresponding electromagnetic structure function. Indeed,
for $F_{2,{\rm CC}}$ which includes both transverse and longitudinal structure functions, the $Q^2\to 0$ limit of the longitudinal structure function $F_L$ does not vanish.
In ref. \cite{Kulagin:2007ju}, Kulagin and Petti find that for $F_{2,{\rm CC}}=(F_T+F_L)/(1+4x^2m_N^2/Q^2)$, PCAC corrections to $F_L$ are
\begin{eqnarray}
    F_L^{\rm PCAC} &=& \frac{f_\pi^2 \sigma_\pi (W^2)}{\pi} f_{\rm PCAC}(Q^2)\\
    \nonumber
    f_{\rm PCAC}(Q^2) &=& \Biggl(1+\frac{Q^2}{M_{\rm PCAC}^2}\Biggr)^{-2}\\
    \nonumber
    \sigma_\pi &\simeq & X (W^2)^\epsilon + Y (W^2)^{-\eta_1}\,.
\end{eqnarray}
Here, $f_\pi=0.93 m_\pi$ is the pion decay constant, $M_{\rm PCAC}= 0.8$ GeV \cite{Kulagin:2007ju}. The parameters in the $\pi p$ cross section $\sigma_\pi$ are $\epsilon\simeq 0.09$, $\eta_1\simeq 0.36$, $X\simeq 12.1$ mb and $Y_1\simeq 26.2$ mb, given $W^2$ in units of GeV. For $Q^2\to 0$ keeping $W^2$ fixed, the longitudinal structure function written in terms of the constants $a$ and $b$ in Table 1 are
\begin{equation}
\label {eq:flpcac}
    F_L(x,Q^2)|_{Q^2=0} \simeq
    A_L \Biggl( \frac{W^2}{a}  \Biggr)^\epsilon + B_L
    \Biggl( \frac{W^2}{b}\Biggr)^{-\eta_1}\,,
\end{equation}
where $A_L\simeq 0.147$ and $B_L\simeq 0.423$.

Precisely how the full $x$ and $Q^2$ dependence of $F_L(x,Q^2)$ proceeds in weak interactions is still to be determined. To estimate how a non-zero limit of $F_{2,{\rm CC}}(x,Q^2)$ for $Q^2\to 0$ would affect the neutrino and antineutrino cross section and to compare the NNSF$\nu$(D) structure function, we parameterize $F_{2,{\rm CC}}^{\rm PCAC}(x,Q^2)$ with 
\begin{eqnarray}
 \nonumber
F_{2,{\rm CC}}&^{\rm PCAC}&= \Biggl[
A^{\rm PCAC}x^{-\Delta(Q^2)}(1-x)^{n(Q^2)+4}\\
\nonumber &\times& \Biggl(
\frac{Q^2}{Q^2+a}\Biggr)^{\Delta(Q^2)}\\ \nonumber
&+& B^{\rm PCAC}x^{1-\alpha_R}(1-x)^{n(Q^2)}\Biggl( \frac{Q^2}{Q^2+b}\Biggr)^{\alpha_R-1}\\ 
&\times & \Bigl( 1+ f(1-x)\Bigr)\Biggr] f_{\rm PCAC}(Q^2) \ 
\label{eq:ckmtpcac}
\end{eqnarray}
with $A^{\rm PCAC} = A_L\simeq 0.147$ and $B^{\rm PCAC}=B_L/(1+f)\simeq 0.265$. This form of $F_{2,{\rm CC}}^{\rm PCAC}$ reduces to eq. (\ref{eq:flpcac}) in the low $Q^2$ limit holding $W^2$ fixed, as long as we can approximate $\epsilon\simeq \Delta_0$ and $\eta_1\simeq 1-\alpha_R$. This is not exact, but it gives a good starting point for understanding how big the PCAC corrections to the neutrino and antineutrino cross sections will be. 
We keep $F_{1,{\rm CC}}$ and $F_{3,{\rm CC}}$ as above. With this approximation, the PCAC correction to $F_{2,{\rm CC}}$ can be almost $\sim 10\%$  for some values of $x$ at $Q^2=2$ GeV$^2$, so we take the normalization scale $Q_0^2=4$ GeV$^2$ where PCAC corrections are smaller, and we make the following substitution for $Q^2<4$ GeV$^2$ to include PCAC corrections. We set
\begin{eqnarray}
\nonumber
F_{2,{\rm CC}}(x,Q^2)
& = &\Biggl[F_{2,{\rm CC}}^{\rm CKMT}(x,Q^2) + F_{2,{\rm CC}}^{\rm PCAC}(x,Q^2)\Biggr]
\\
 &\times & 
 \frac{F_{2,{\rm CC}}^{\rm NLO+TMC}(x,Q_0^2)}{F_{2,{\rm CC}}^{\rm CKMT}(x,Q_0^2)}\,,
 \label{eq:pcacnorm}
\end{eqnarray}
and we designate the structure function normalized this way as CKMT+PCAC-NT.

\begin{figure} [t]
\centering
\includegraphics[width=.47\textwidth]{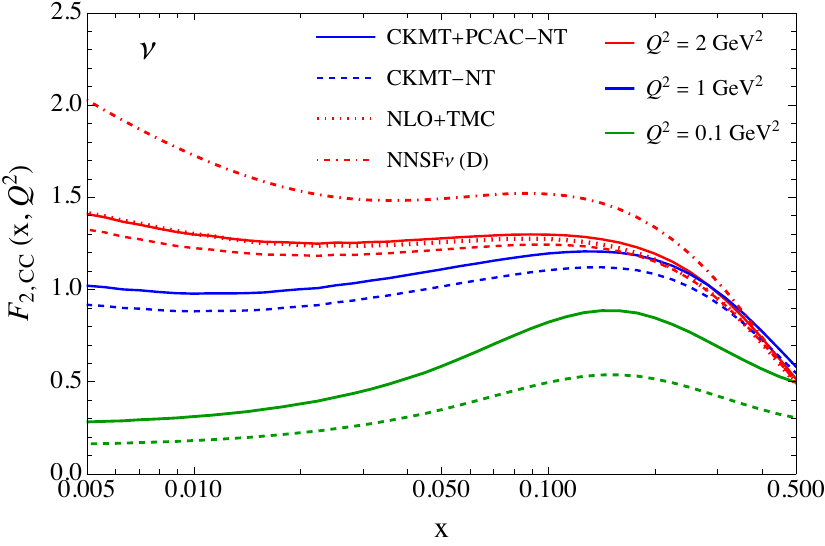}
\includegraphics[width=.47\textwidth]{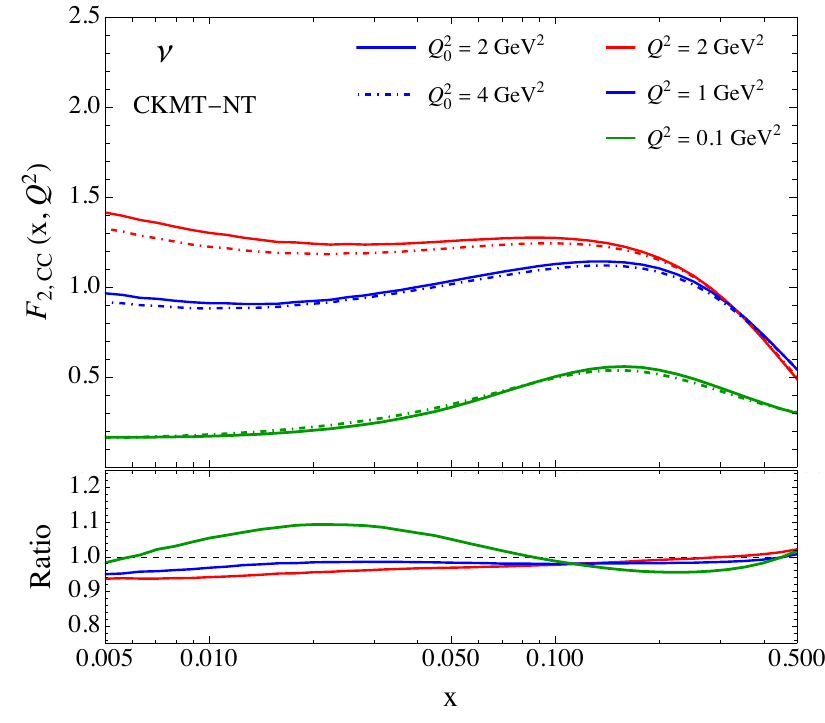}
\caption{\label{Fig:F2pcac} 
Upper: The structure functions $F_{2,{\rm CC}} (x, Q^2)$ for neutrino-nucleon charged current scattering for $Q^2$ = 0.1, 1 and 2 $\rm GeV^2$ with the CKMT-NT form of the structure function (dashed) extrapolated below $Q_0^2$ normalized according to eq. (\ref{eq:norm}) and the CKMT+PCAC-NT (solid) normalized according to eq. (\ref{eq:pcacnorm}), both for $Q_0^2=4$ GeV$^2$. For $Q^2=2$ GeV$^2$, the structure function for 
NNSF$\nu$(D)  (dot-dashed) \cite{Candido:2023utz} and 
for NLO+TMC (dotted) are presented for comparison. 
Lower: The structure functions $F_{2,{\rm CC}} (x, Q^2)$ with CKMT-NT for two normalization scales, $Q_0^2$ = 2 and 4 GeV$^2$, and the the ratio of the CKMT-NT results with  $Q_0^2$ = 4 GeV$^2$ to $Q_0^2$ = 2 GeV$^2$, for $Q^2$ = 0.1, 1 and 2 $\rm GeV^2$.
}
\end{figure}

Figure \ref{Fig:F2pcac} shows a comparison of $F_{2,{\rm CC}} (x, Q^2)$
for normalized CKMT (eq. (\ref{eq:norm}), CKMT-NT) and normalized CKMT+PCAC (eq. (\ref{eq:pcacnorm}), CKMT+PCAC-NT), both with $Q_0^2=4$ GeV$^2$, for $Q^2$ = 0.1, 1 and 2 $\rm GeV^2$. Also shown for reference are the NLO+TMC and NNSF$\nu$(D) structure functions for $Q^2=2$ GeV$^2$. Even with the PCAC correction, $F_{2,{\rm CC}} (x, Q^2)$ in our approach (CKMT+PCAC-NT) is much smaller than the NNSF$\nu$(D) structure function for most of the range of $x$.

\section{Neutrino cross sections}
\label{sec:xc}

In this section, the impacts of $W$ and $Q^2$ kinematic regions on the neutrino DIS CC cross sections are investigated using both the CKMT-NT and CKMT+PCAC-NT extrapolations. All cross sections are for isoscalar nucleon targets unless specified.
We evaluate the $\nu_\mu$, $\bar\nu_\mu$, $\nu_\tau$ and $\bar\nu_\tau$ CC cross sections with a default value of hadronic final state invariant mass $W>W_{\rm min}=1.4$ GeV. 

\begin{figure} 
\centering
\includegraphics[width=.47\textwidth]{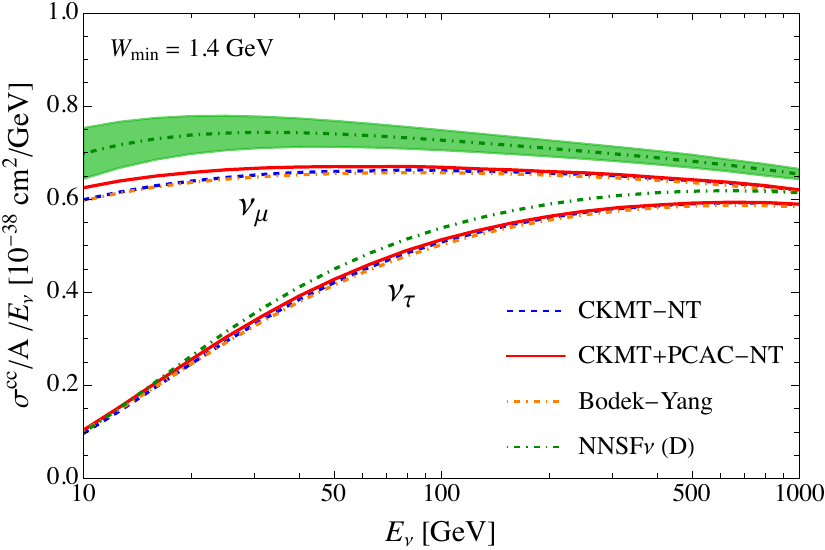}
\includegraphics[width=.47\textwidth]{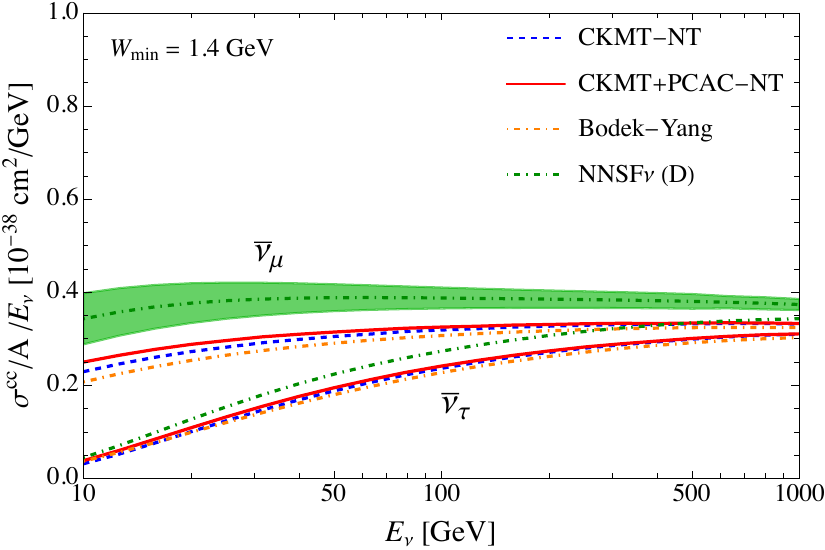}
\caption{\label{Fig:CSnuA} 
The charged current cross sections per nucleon of muon neutrino and tau neutrino (upper) and their antineutrino (lower) scattering with isoscalar nucleons for $W_{\rm min}$ = 1.4 GeV for CKMT-NT and CKMT+PCAC-NT evaluations. The predictions evaluated with the Bodek-Yang and NNSF$\nu$ with its error band are presented for comparison. 
}
\end{figure}

Results for $\sigma_{\rm CC}/E_\nu$ are shown in Fig. \ref{Fig:CSnuA} for $\nu_\mu N$ and $\nu_\tau N$ CC scattering (upper panel) and $\bar{\nu}_\mu N$ and $\bar\nu_\tau N$ CC scattering (lower panel). 
Cross sections for CKMT-NT evaluated with $Q_0^2=4\ \GeV2$ agree with cross sections  evaluated with $Q_0^2=2$ GeV$^2$ to within $1\%$ for $\nu_\mu N$ CC scattering and to within $2\%$ for $\bar\nu_\mu N$ CC scattering, with similar results for $\nu_\tau N$ and $\bar\nu_\tau N$ CC scattering with $E_\nu \gtrsim 10$~GeV.

Also shown in Fig. \ref{Fig:CSnuA} are the cross sections evaluated using the Bodek-Yang model using GRV98LO PDFs (dot-dashed) and with the NNSF$\nu$(D) structure functions. There is very good agreement for $\sigma_{\rm CC}/E_\nu$ between the Bodek-Yang and CKMT-NT evaluations. 
The CKMT+PCAC-NT cross sections are larger than the CKMT-NT evaluations by $\sim 4 (7)\%$ for $\nu N$ scattering and $\sim 10 (20)\%$ for $\bar{\nu} N$ scattering at $E_\nu=10$ GeV for muon neutrinos (tau neutrinos).
These cross sections are closer to each other at higher energies, where larger $Q^2$ values contribute more to the cross sections.

Given that the NNSF$\nu$(D) structure functions shown in Fig. \ref{Fig:F2} are higher than the other curves, it is not surprising that the NNSF$\nu$(D) cross sections are also larger than the other evaluations in Fig. \ref{Fig:CSnuA}. 
As noted in ref. \cite{Candido:2023utz}, the NNSF$\nu$(D) uncertainty bands for the CC cross section do not overlap the Bodek-Yang cross section for $\nu_\mu N$ scattering for $W>2$ GeV. 
This is also the case for $W>1.4$ GeV. As shown in Fig.~\ref{Fig:CSnuA}, the NNSF$\nu$(D) uncertainty bands overlap neither our cross sections with the CKMT+PCAC-NT nor the Bodek-Yang cross sections.

\begin{figure}
\centering
\includegraphics[width=.47\textwidth]{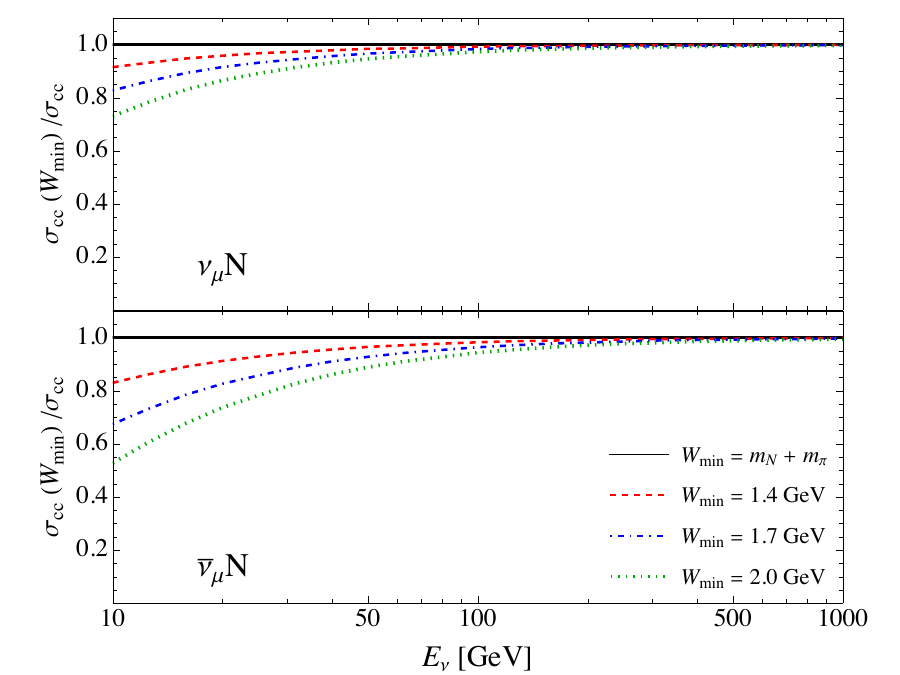}
\includegraphics[width=.47\textwidth]{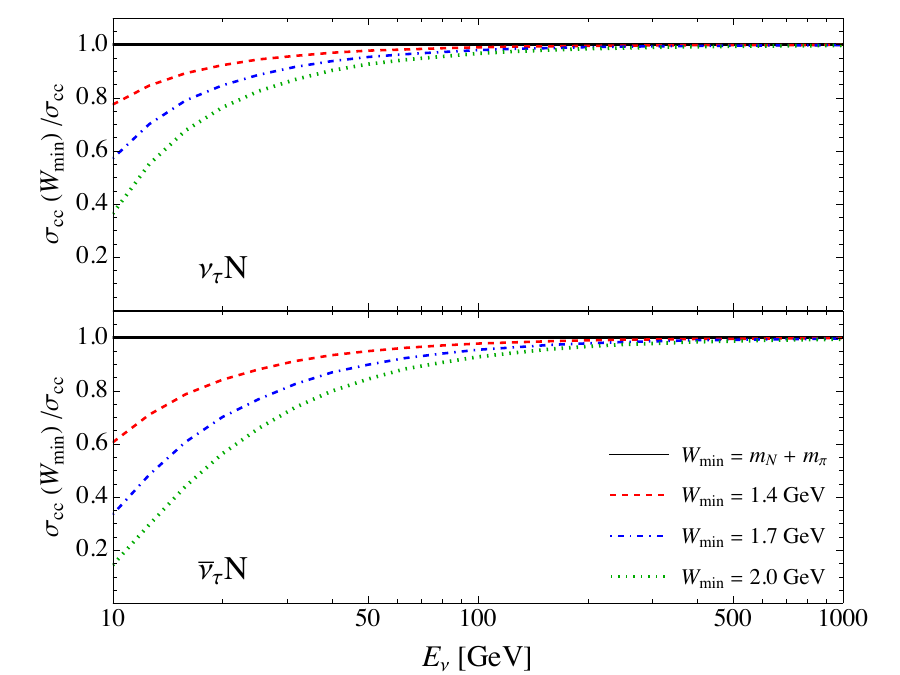}
\caption{\label{Fig:RCSnuN-Wmin} 
The ratios of the charged current cross sections for muon neutrino and antineutrino (upper) and tau neutrino and antineutrino (lower) scattering with isoscalar nucleons as function of energy for the values of the minimum hadronic final state invariant mass $W_{\rm min}$ = 1.4, 1.7 and 2 GeV to the cross sections with $W_{\rm min} = m_N + m_\pi$ using CKMT+PCAC-NT with $Q_0^2=4$ GeV$^2$. 
}  
\end{figure}

The remaining figures in this section come from evaluations with CKMT+PCAC-NT with $Q_0^2=4$ GeV$^2$. Fig.~\ref{Fig:RCSnuN-Wmin} shows the ratios of charged current cross sections for $W_{\rm min}$ = 1.4 GeV, 1.7 GeV and 2.0 GeV with respect to the cross section with $W_{\rm min}= m_{N} + m_\pi$, the minimum hadronic invariant mass for inelastic scattering. The figure shows the results for muon neutrinos and antineutrinos in the upper panel, and for tau neutrinos and antineutrinos in the lower panel. 
The effect of $W_{\rm min}$ appears at energies below a few hundred GeV, and it is more significant at lower energies.
The cuts on $W_{\rm min}$ reduce more the cross sections for tau neutrinos than the cross sections for muon neutrinos for both neutrinos and antineutrinos.
While the cross sections with $W_{\rm min} = 2.0~{\rm GeV}$, for example, are suppressed by 3\% compared to the results with $W_{\rm min}= m_{N} + m_\pi$ at $E_\nu = 100~{\rm GeV}$ for both muon neutrinos and tau neutrinos, the suppression at 10 GeV is approximately 27\% for muon neutrinos and 64\% for tau neutrinos. 
In addition, the impact of $W_{\rm min}$ is larger on the antineutrino cross sections than the neutrino cross sections.
The corresponding suppression for $\bar{\nu}_\mu$ and $\bar{\nu}_\tau$ CC cross sections is 6 -- 7\% at 100 GeV, and about 47\% and 86\% at 10 GeV, respectively. 
Given the fact that for $\Delta (1232)$ resonance production, $1.1~{\rm GeV} \lesssim W \lesssim 1.4~{\rm GeV}$, one can use $W>W_{\rm min}=1.4$ GeV to exclude the $\Delta$ resonance region. For $W>W_{\rm min}=1.4$ GeV, the cross sections are suppressed 
by $\sim 10$ (15)\% and $\sim 25$ (40)\% with respect to the results with $W_{\rm min}= m_{N} + m_\pi$ for $\nu_\mu$ and $\nu_\tau$ (antineutrinos) at $E_\nu = 10~{\rm GeV}$, and such impacts are reduced to less than $\sim 2\%$ at $E_\nu = 100~{\rm GeV}$. 
The dependence of the cross sections on $W_{\rm min}$ is not very sensitive to whether or not $F_{2,{\rm CC}}^{\rm PCAC}$ is included.

\begin{figure}
\centering
\includegraphics[width=.47\textwidth]{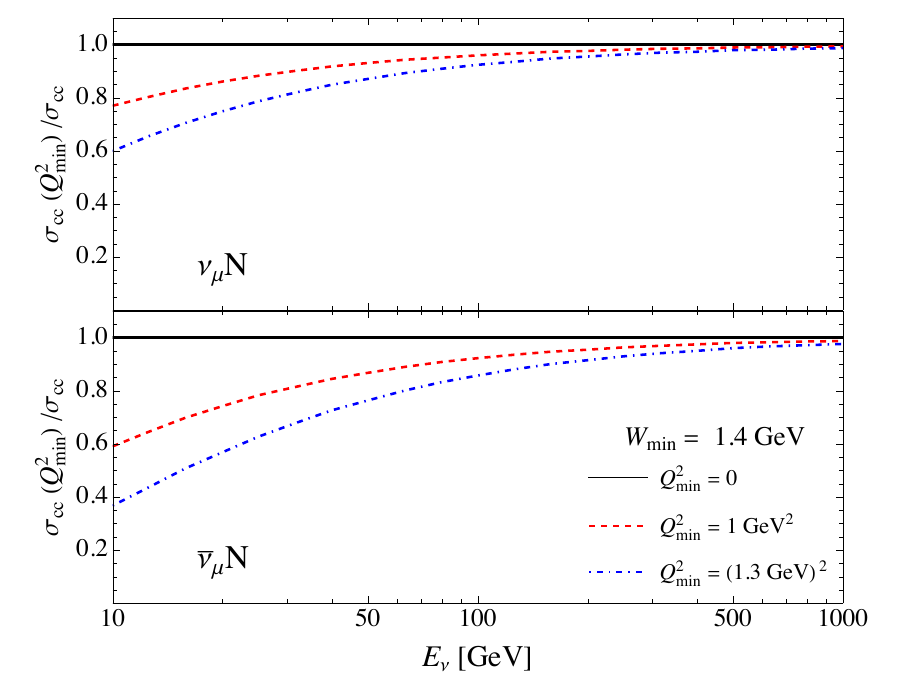}
\includegraphics[width=.47\textwidth]{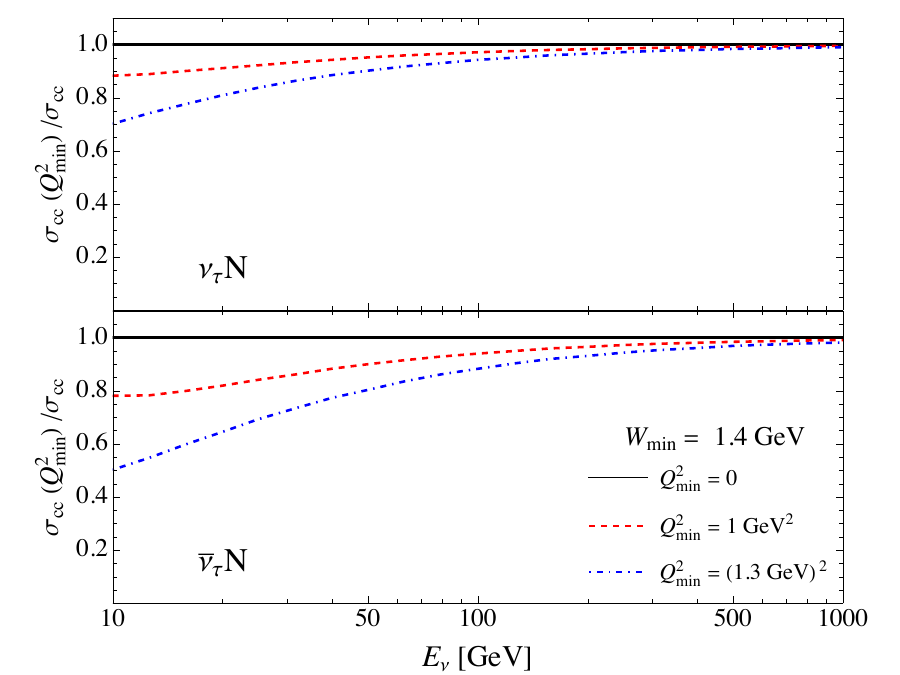}
\caption{\label{Fig:RCSnuN-Q2min} 
The ratio of the CC cross sections with the cut of $Q_{\rm min}^2$ = 1 and $1.3^2$ GeV$^2$ to the ones with full $Q^2$ range for the scattering of muon neutrino and antineutrino (upper), and tau neutrino and antineutrino (lower) with nucleon for $W_{\rm min}= 1.4~{\rm GeV}$
using CKMT+PCAC-NT with $Q_0^2=4$ GeV$^2$. 
}  
\end{figure}

The $Q^2$ dependence of the neutrino and antineutrino CC cross sections are shown in Fig.~\ref{Fig:RCSnuN-Q2min}. 
In the figure, the ratios of the cross sections evaluated with the minimum values of $Q^2_{\rm min}$ = 1 GeV$^2$ and $1.69~{\rm GeV}^2$ to those with full $Q^2$ range  are presented for muon neutrinos (upper) and tau neutrinos (lower) with $W_{\rm min}=1.4$ GeV as in Fig.~\ref{Fig:CSnuA}. 
Considering eq. (\ref{eq:w2}),
$W>W_{\rm min}$ enforces a minimum  $Q^2$ with a  value greater than 0 as a function of parton momentum fraction $x$.
Nonetheless, we denote $Q_{\rm min}^2=0$ to indicate that there are no additional restrictions on $Q^2$ beyond the requirement that $W>W_{\rm min}$.
The value of 1.3 GeV for $Q$ is selected from the minimum value of the nCTEQ15 PDFs used in our evaluation.

From the upper panel of the figure for muon neutrinos (upper) and antineutrinos (lower), one can see that the impact of $Q^2_{\rm min}$ appears over a wider energy range with more significant effects than the impacts of $W_{\rm min}$. 
The figure indicates that for example, at $E_\nu = 10~{\rm GeV}$, 40\% and 63\% of the muon neutrino and antineutrino cross sections come from the region of $Q^2 \lesssim 1.69~{\rm GeV}^2$. 
Such large effect is reduced with increasing energy, however, even for $E_\nu$ = 100 GeV, the corresponding contributions to the $\nu_\mu N$ and $\bar\nu_\mu N$ cross sections from $Q^2<Q^2_{\rm min} \simeq 1.69~{\rm GeV^2}$ are about 8\% and 14\%, respectively.

\begin{figure*}
\centering
\includegraphics[width=.47\textwidth]{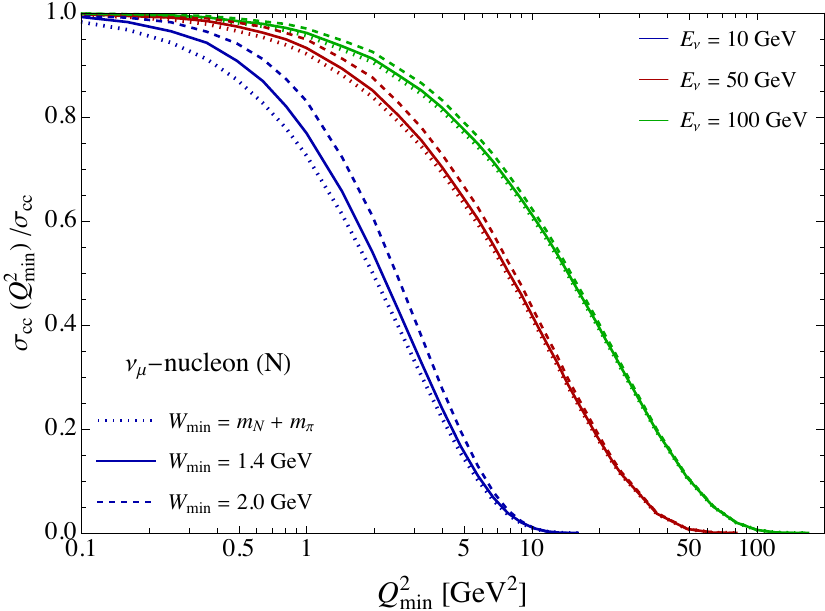}
\includegraphics[width=.47\textwidth]{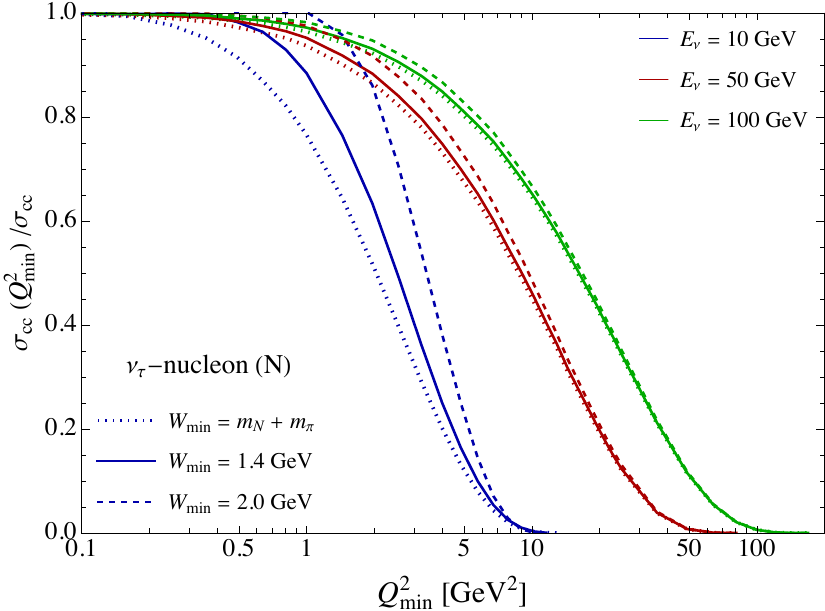}
\includegraphics[width=.47\textwidth]{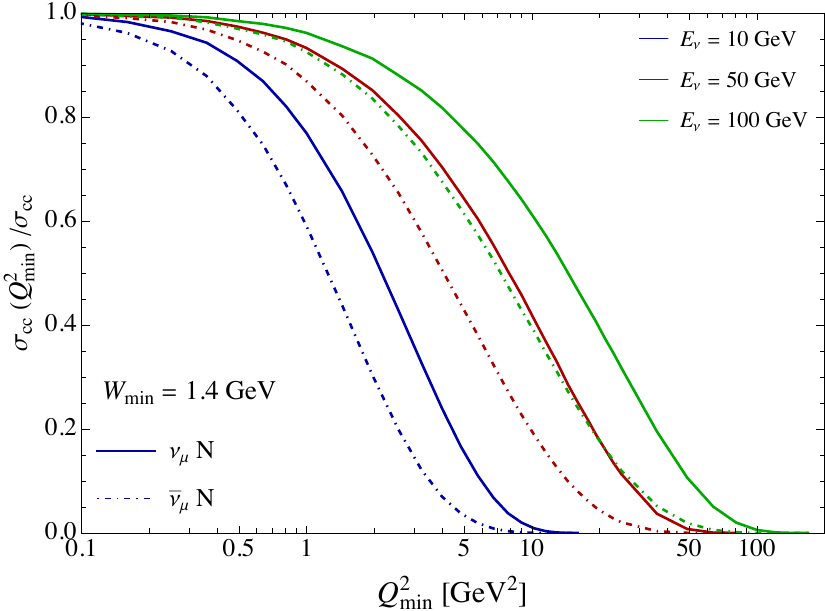}
\includegraphics[width=.47\textwidth]{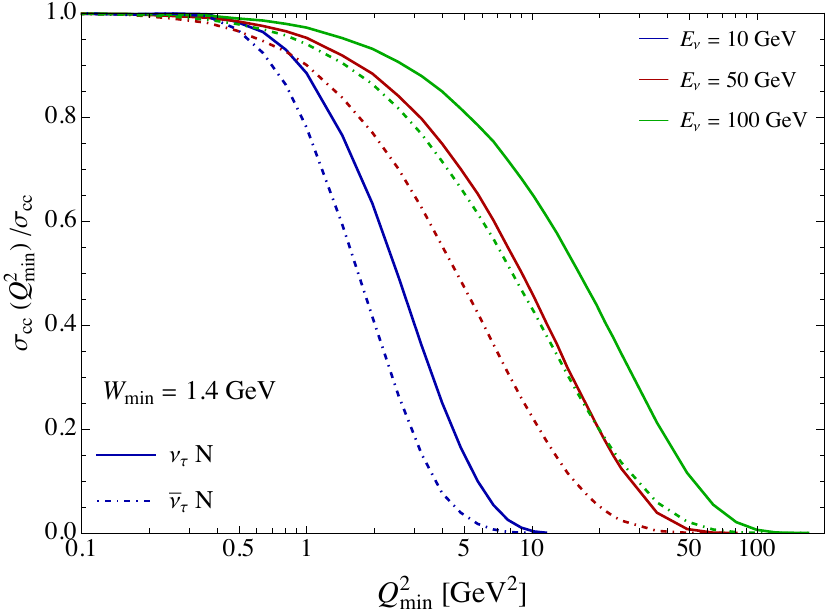}
\caption{\label{fig:RCSnuN-Qmin2}
The ratio of the CC neutrino-isoscalar nucleon cross sections with a minimum value of $Q^2$ to the cross section with all $Q^2$, as function of $Q^2_{\rm min}$. The upper row shows the results for muon neutrinos (left) and tau neutrinos (right) for the $W_{\rm min}=m_N + m_\pi$, 1.4 and 2 GeV, whereas the results for neutrino and antineutrino for $W_{\rm min}$ = 1.4 GeV are presented in the lower panels. Here, we use
CKMT+PCAC-NT with $Q_0^2=4$ GeV$^2$.
}
\end{figure*}

For tau neutrino scattering cross sections, the contributions from $Q^2 \leq 1.69~{\rm GeV^2}$ are comparable to the muon neutrino case at $E_\nu$ = 100 GeV. 
On the other hand, the respective fractions of the cross sections at 10 GeV 
become 30\% and 50\%, which shows the impact of the cutoff on $Q^2$ is lower than the impact on the muon neutrino and antineutrino cross sections when $W_{\rm min} = 1.4~{\rm GeV}$. 
This lower impact of $Q^2$ for tau neutrinos and antineutrinos at low energies reflects kinematic constraints on tau production given $W_{\rm min}$ \cite{Reno:2021hrj}, as also shown in Fig.~\ref{fig:RCSnuN-Qmin2}.

Figure~\ref{fig:RCSnuN-Qmin2} presents the ratios as in Fig.~\ref{Fig:RCSnuN-Q2min} in more detail for selected energies. The CC cross sections normalized to the cross sections with full $Q^2$ range are shown, now as function of $Q^2_{\rm min}$. 
We show the results for several values of $W_{\rm min}$ and for $E_\nu$ = 10, 50 and 100 GeV for muon neutrinos in the left panels and for tau neutrinos in the right panels.
From the upper panels, one can see that the impact of kinematic restrictions appears more clearly at lower energies.
First, the results with the different values of $W_{\rm min}$ are distinguishable for $E_\nu = 10~{\rm GeV}$ while the corresponding results for $E_\nu = 100~{\rm GeV}$ almost overlap.
The effect of the cutoff on $Q^2$ is more appreciable.  
For muon neutrinos, for example, setting $Q^2_{\rm min} = 1~{\rm GeV^2}$ suppresses the cross section by 23\% at 10 GeV whereas it reduces only 4\% at 100 GeV when $W_{\rm min}$ = 1.4 GeV.
For tau neutrinos, the fractions with the same conditions are 12\% and 3\% at 10 GeV and 100 GeV, respectively. 
In case of tau neutrinos, the influence of kinematic conditions for tau production and $W_{\rm min}$ is included as shown in the low energy results obviously in particular for 10 GeV. 
As shown in Fig.~\ref{Fig:RCSnuN-Wmin}, the restrictions for tau production and the cutoff on $W_{\rm min}$ already significantly suppress the tau neutrino cross sections with full $Q^2$ range, and consequently the low $Q^2$ impact on the ratios diminishes as $W_{\rm min}$ increases. 
In the lower panels of Fig.~\ref{fig:RCSnuN-Qmin2}, the ratios of the CC cross sections for antineutrinos are compared with the results for neutrinos for $W_{\rm min}$ = 1.4 GeV. 
It is evident that the cutoff on the $Q^2_{\rm min}$ has more significant impact on the antineutrino cross sections than on the neutrino cross sections for both muon neutrinos and tau neutrinos.
In addition, we evaluated these ratios with different PDF sets,  HERAPDF20 \cite{H1:2015ubc}, NNPDF4.0 \cite{NNPDF:2021njg}, MSHT \cite{Bailey:2020ooq}, and checked that the results are largely independent of the PDF choice. 
For $\sigma_{cc} (Q^2_{\rm min}) / \sigma_{cc} \gtrsim$ 0.1, the ratios with the different PDFs differ by less than 10\% from the one with our default PDFs, nCTEQ15.

\section{Discussion}
\label{sec:conclusions}

We have investigated contributions of the $W$ and $Q^2$ outside the DIS region to the inelastic CC cross sections of neutrinos and antineutrinos incident on isoscalar targets, using theoretically motivated and phenomenological parameterizations of the low-$Q^2$ weak structure functions as a starting point to developing a better understanding of this interesting kinematic regime.
As mentioned in section \ref{sec:SF}, the DIS region is generally defined as $W>2$ GeV and $Q^2 > 1~{\rm GeV^2}$, whereas in GENIE~\cite{GENIE:2021zuu}, $W_{\rm min}= 1.7$ GeV is used for the DIS contribution to the cross section. Our interest in the low $W$ and low $Q^2$ region is in part motivated by current and up-coming experiments to detect neutrinos produced in the forward direction from LHC collisions with neutrino energies up to a few TeV.

We have focused here on the relative sizes of contributions of different kinematic regions to the neutrino-nucleon CC cross sections. In our evaluations using the CKMT extrapolation of $F_{2,{\rm EM}}$ ~\cite{Capella:1994cr,Kaidalov:1998pn} adapted to $F_{2,{\rm CC}}$ \cite{Reno:2006hj} and matched to NLO QCD with target mass corrections, together with an approximation to the PCAC correction \cite{Kulagin:2007ju}, the contributions to the CC neutrino (antineutrino) cross sections from the range of $m_N+m_\pi<W<2~{\rm GeV}$ are less than $\sim 3(7)\%$  at $E_\nu=100$ GeV for both muon neutrinos and tau neutrinos, and they are even less at higher energies. The low-$Q^2$ region contributes a small fraction of neutrino CC cross sections for $E_\nu>100$ GeV.
For $W_{\rm min}=2~{\rm GeV}$,  3(6)\% of the $\nu_\mu (\bar\nu_\mu)$ CC cross sections comes from $Q^2 < 1~{\rm GeV^2}$ at $E_\nu = 100~{\rm GeV}$. 
This is comparable to results with $W_{\rm min}=1.4~{\rm GeV}$ as can be seen in Fig.~\ref{fig:RCSnuN-Qmin2}. 
Slightly lower CC cross section fractions come from $Q^2<1$ GeV$^2$ for $\nu_\tau$ and $\bar{\nu}_\tau$. 
Thus, our conclusion is that the impact of low $W$ and $Q^2$ outside the range for DIS interaction on the neutrino and antineutrino cross sections is at most a few percent level for a neutrino energy of 100 GeV, 
and negligible for neutrino energies above a few hundred GeV, the main energy range that can be explored by the FPF.

\begin{figure} [t]
\centering
\includegraphics[width=.47\textwidth]{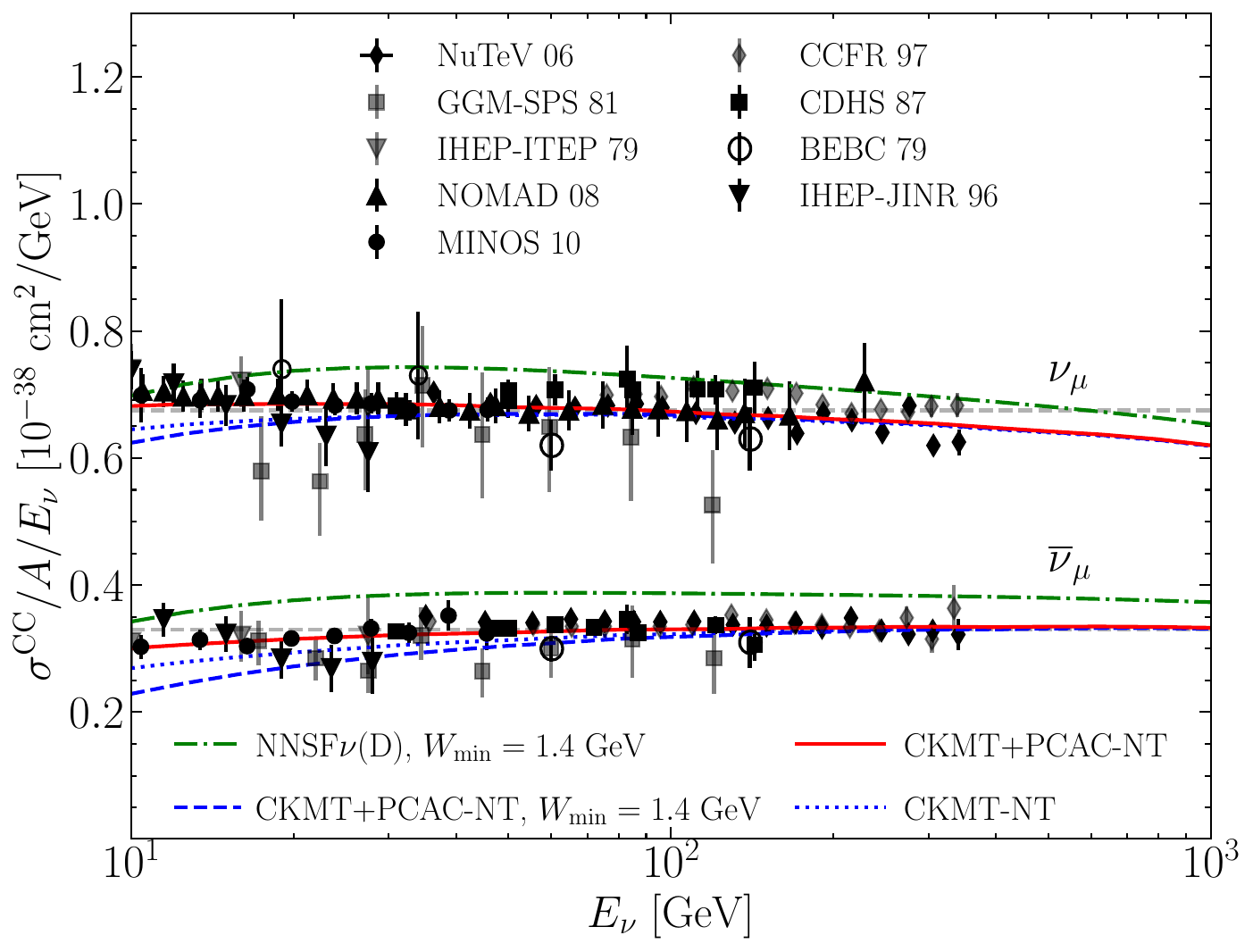}
\caption{\label{Fig:xcwithdata} 
The $\nu_\mu N$ and $\bar\nu_\mu N$ CC cross sections for $W_{\rm min}=1.4$ GeV using CKMT+PCAC-NT (dashed) and NNSF$\nu$(D) structure functions for deuterium (dot-dashed), and for $W_{\rm min}=m_N+m_\pi$ for CKMT-NT  (dotted) and CKMT+PCAC-NT normalized according to eq. (\ref{eq:pcacnorm}) (solid). Data are shown for neutrino experiments using a range of targets \cite{NuTeV:2005wsg,GargamelleSPS:1981hpd,Mukhin:1979bd,NOMAD:2007krq,MINOS:2009ugl,Seligman:1997fe,Berge:1987zw,deGroot:1978feq,Anikeev:1995dj}. The light gray dashed lines are at 0.675 and 0.33. }
\end{figure}

Our CC cross sections differ from predictions using NNSF$\nu$(D) structure functions \cite{Candido:2023utz}.  Figure \ref{Fig:xcwithdata} shows $\sigma^{\rm CC}/A/E_\nu$ with theoretical evaluations for  $\nu_\mu$ and $\bar\nu_\mu$ CC scattering on isoscalar nucleons and data from neutrino experiments with a range of targets \cite{NuTeV:2005wsg,GargamelleSPS:1981hpd,Mukhin:1979bd,NOMAD:2007krq,MINOS:2009ugl,Seligman:1997fe,Berge:1987zw,deGroot:1978feq,Anikeev:1995dj}.
The dot-dashed curve shows the NNSF$\nu$(D) cross section for $W_{\min}=1.4$ GeV and the dashed curve shows our CKMT+PCAC-NT result, also with $W_{\min}=1.4$~GeV.
For reference, we also show the CKMT-NT and CKMT+PCAC-NT for $W_{\min}=m_N+m_\pi$, where the latter (solid curve) represents our best approximation of the full CC cross section even at $E_\nu=10$ GeV.
For $E_\nu=200$~GeV where $W_{\rm min}$ is not important, the NNSF$\nu$(D) cross section is a factor of 
$\sim$ 1.1 larger for $\nu_\mu N$ CC scattering and a factor of 
$\sim1.2$ larger for $\bar\nu_\mu N$ CC scattering than the CKMT+PCAC-NT result. Cross sections evaluated using the Bodek-Yang model agree well with the CKMT+PCAC-NT evaluations, as shown in Fig. \ref{Fig:CSnuA}.

\begin{figure} [h]
\centering
\includegraphics[width=.47\textwidth]{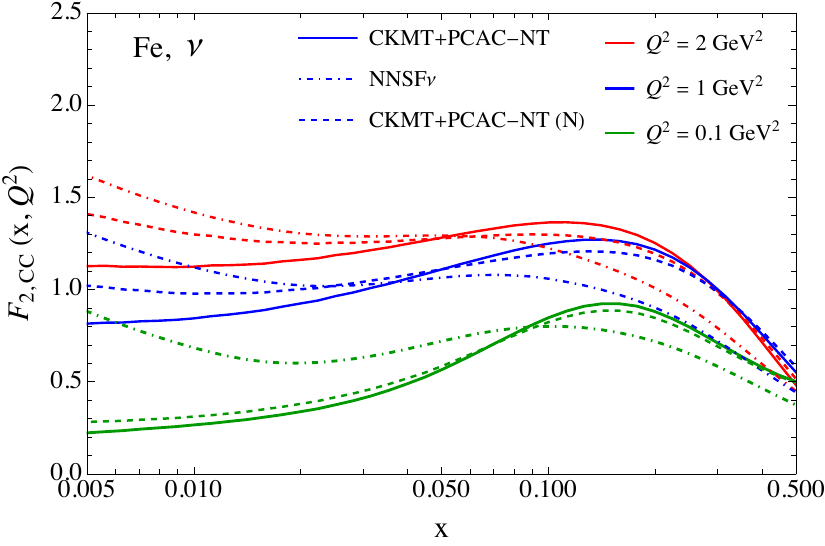}
\includegraphics[width=.47\textwidth]{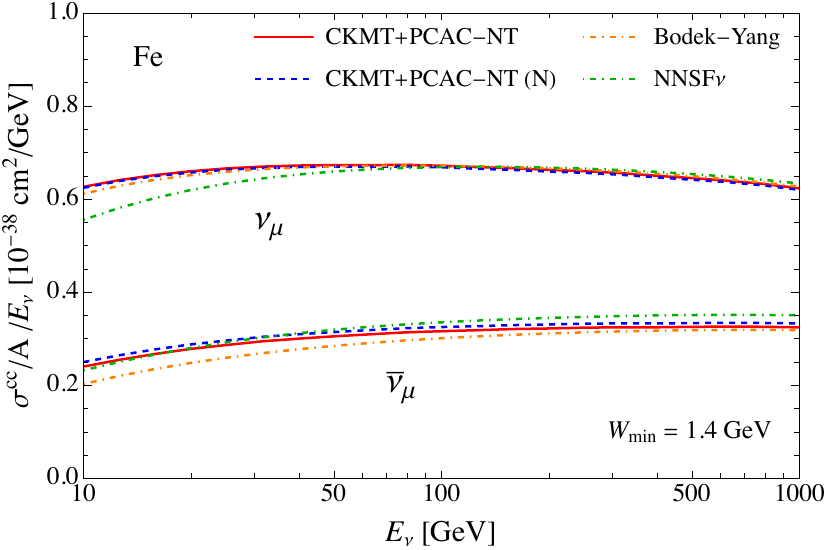}
\caption{\label{Fig:xcFe} 
Upper: the structure functions $F_{2,{\rm CC}} (x, Q^2)$ for neutrino CC scattering with iron (Fe) for $Q^2$ = 0.1, 1 and 2 $\rm GeV^2$ with the CKMT+PCAC-NT (solid) and NNSF$\nu$(Fe) (dot-dashed).
Lower: the $\nu_\mu N$ and $\bar\nu_\mu N$ CC cross sections for $W_{\rm min}=1.4$ GeV using NNSF$\nu$(Fe) structure functions (dot-dashed) and CKMT+PCAC-NT using nCTEQ15 iron PDFs, normalized according to eq. (\ref{eq:pcacnorm}) (solid). Also shown are the Bodek-Yang cross sections for comparison.
}
\end{figure}

\begin{figure} 
\centering
\includegraphics[width=.47\textwidth]{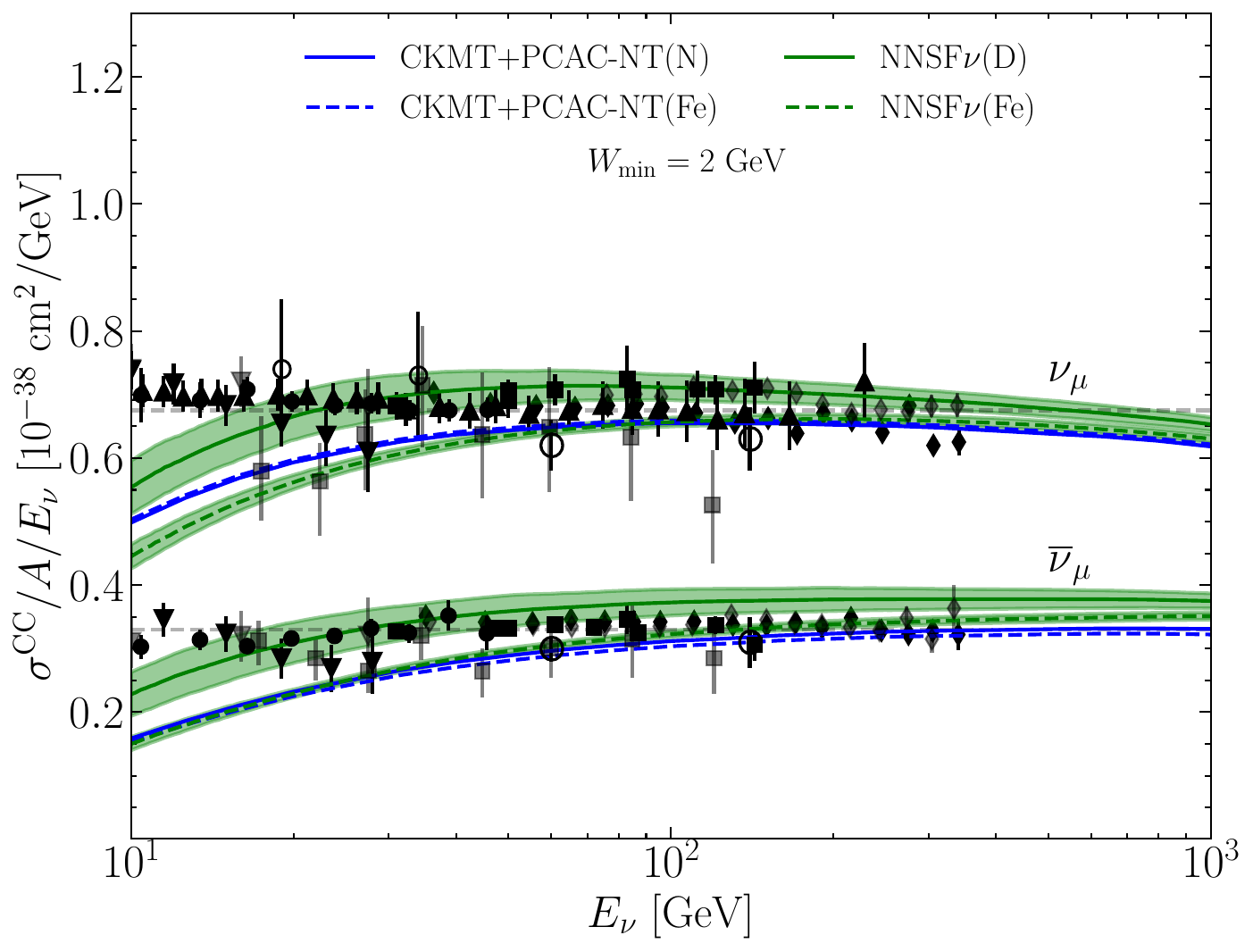}
\caption{\label{Fig:xcwithdata-wmin2} 
A comparison of the $\nu_\mu A$ and  $\bar{\nu}_\mu A $ CC cross sections  divided by incident energy per $A$  for
$W_{\rm min} = 2$ GeV for $A=2 $ (isoscalar nucleons, solid) and $A=56$ (Fe, dashed)  for CKMT+PCAT-NT and
NNSF$\nu$ structure functions. The results with NNSF$\nu$ are shown with uncertainty bands. The data are shown as in Fig. \ref{Fig:xcwithdata}.}
\end{figure}

As noted, the data shown in Fig. \ref{Fig:xcwithdata} come from neutrino experiments using a range of targets. For example, the data for energies up to $\sim 350$ GeV from
CCFR  \cite{Seligman:1997fe} and NuTeV \cite{NuTeV:2005wsg} come from neutrino and antineutrino scattering on iron. The NNSF$\nu$(Fe) structure functions are more similar to the CKMT+PCAC-NT structure functions evaluated with nCTEQ15 iron PDFs than the NNSF$\nu$(D) structure functions  are to the CKMT+PCAC-NT isoscalar nucleon structure functions. 
The upper panel of Fig. \ref{Fig:xcFe} shows $F_{2,{\rm CC}}$ for iron, and the lower panel shows the cross sections per nucleon, scaled by incident energy, for $\nu_\mu$ and $\bar\nu_\mu$ CC scattering given $W_{\rm min}=1.4$ GeV.
The dashed curves show the same quantities for CKMT+PCAC-NT from scattering with isoscalar nucleons. The agreement between 
the evaluations using CKMT+PCAC-NT and NNSF$\nu$(Fe) for iron targets is better than for isoscalar nucleon targets.

A similar agreement between CKMT+PCAC-NT and NNSF$\nu$(Fe) for iron targets is seen in Fig. \ref{Fig:xcwithdata-wmin2} where now $W_{\rm min}=2$ GeV. 
This is the minimum $W$ for the NNSF$\nu$  structure function analysis. The NNSF$\nu$ error bands for isoscalar and iron targets are also shown. 

The disagreements between isoscalar nucleon results for CKMT+PCAC-NT and NNSF$\nu$(D) CC cross sections persist for $W_{\rm min}=2$ GeV, even at high energies where one expects that the PDFs rather than the low-$Q^2$ structure function extrapolations account for most of the cross section. 
In our approach, PDF uncertainties in the CC cross section cannot account for the discrepancies. For example, for $E_\nu=100$ GeV using the default PDFs,  PDF uncertainties are $\sim 2\%$ for $\nu_\mu N$ and $\sim 4\%$ for $\bar\nu_\mu N$ CC cross sections with $W_{\rm min}=2$ GeV.  
We also compared with the results evaluated using the central set of other PDFs, HERAPDF20 \cite{H1:2015ubc}, NNPDF4.0 \cite{NNPDF:2021njg}, MSHT \cite{Bailey:2020ooq}, which agree with our results with default PDFs, nCTEQ15 for the CKMT+PCAC-NT, to within $\pm$ 4\% for both muon neutrinos and antineutrinos for 10~GeV $ \leq E_\nu \leq 1$~TeV.

The fits of NNSF$\nu$ to neutrino and antineutrino data are from scattering with Ne, CaCO$_3$, Fe and Pb targets. The good agreement between results for scattering with Fe 
may suggest that the NNSF$\nu$ extrapolations to $A=2$ have larger uncertainties than described in ref. \cite{Candido:2023utz}. One test of the NNSF extrapolation to low $A$ could be made by performing an extraction of the electromagnetic structure function (NNSF$\gamma$) for $A$ values comparable to those used for NNSF$\nu$, then to compare its extrapolation to smaller $A$ with small $A$ electromagnetic scattering data. In any case, the CC cross section comparisons presented here emphasize the value of new neutrino and antineutrino scattering data in the 10 GeV -- 1 TeV energy range.

In summary, using our approach to extrapolate PDF-based structure functions to low-$Q^2$, for neutrino energies less than 100 GeV, the kinematic regions of $W$ and $Q^2$ are more important than for higher incident neutrino energies, the impact of which can be more than 50\% for $E_\nu = {\cal O} (10) $ GeV.
As indicated above, the low-$Q^2$ structure functions used in this work are theoretically motivated, tied to the electromagnetic structure functions and the partial conservation of the axial vector current, but more work studying the low-$Q^2$ behavior of weak structure functions is merited and has begun (e.g., in ref. \cite{Bodek:2021bde}).

On the experimental side, neutrino experiments at the FPF will be able to measure a large number of neutrinos up to a few TeV energies. 
The majority of the neutrino events will be distributed in the energies above 100 GeV. An estimate of the number of interacting $\nu_\mu+\bar\nu_\mu$ for the FLArE detector at the FPF, with 3000 fb$^{-1}$, is of order $1.5\times 10^5$ for $E_\nu=100$ GeV$-1$ TeV \cite{Kling:personalcommunication}.
However, simulations show that there can be a considerable number of events at lower energies. For example, again for FLArE at the FPF, of order $2\times 10^4$ $\nu_\mu+\bar\nu_\mu$ interactions are anticipated for neutrino energies in the range of $50-100$ GeV \cite{Kling:personalcommunication}. 
An estimate of 2,000 FLArE CC events in the resonance region, and an even larger number of CC events in FASER$\nu$2, will be relevant for the SIS interactions (see section 7.3 of ref.~\cite{Feng:2022inv} and ref. \cite{Batell:2021aja}). From a combined analysis of the neutrino fluxes and cross sections at the FPF \cite{Kling:2021gos}, cross section measurements will provide a unique opportunities to investigate neutrino interactions in the kinematic region for SIS and DIS and at low $Q^2$, complementary to the cross section analyses of DUNE in the future.

\begin{acknowledgments}
This work is supported in part by US Department of Energy grants DE-SC-0010113 (MHR) and the National Research Foundation of Korea (NRF) grant funded by the Korea government through Ministry of Science and ICT grant 2021R1A2C1009296 (YSJ). We thank F. Kling and A. Garcia for discussions, F. Kling and M. Bustamante for providing the data files for the experiments in Fig. \ref{Fig:xcwithdata}, and A. Garcia for providing cross sections for Fig. \ref{Fig:xcwithdata-wmin2}. For facilitating portions of this research, MHR wishes to acknowledge the Center for Theoretical Underground Physics and Related Areas (CETUP*), The Institute for Underground Science at Sanford Underground Research Facility (SURF), and the South Dakota Science and Technology Authority for hospitality and financial support, as well as for providing a stimulating environment.
\end{acknowledgments}

\appendix

\section{Deep inelastic scattering cross section}

We summarize the formulas for neutrino $\nu_\ell (k)+ N(p)\to \ell(k')+X$
and antineutrino 
$\bar\nu_\ell (k)+ N(p)\to \bar\ell(k')+X$ charged-current (CC) cross sections 
in this appendix.
The differential cross sections for neutrino and antineutrino CC scattering can be written
\begin{eqnarray} \nonumber
\frac{d^2\sigma^{\nu(\bar{\nu})}}{dx\ dy} &=& \frac{G_F^2 m_N
E_{\nu}}{\pi(1+Q^2/M_W^2)^2}
\Biggl(
(y^2 x + \frac{m_{\ell}^2 y}{2 E_{\nu} m_N})
F_{1,{\rm CC}} \\ \nonumber
&+& \left[ (1-\frac{m_{\ell}^2}{4 E_{\nu}^2})
-(1+\frac{m_N x}{2 E_{\nu}}) y\right]
F_{2,{\rm CC}}
\\ \nonumber
&\pm& 
\left[x y (1-\frac{y}{2})-\frac{m_{\ell}^2 y}{4 E_{\nu} m_N}\right]
F_{3,{\rm CC}} \\  
&+& 
\frac{m_{\ell}^2(m_{\ell}^2+Q^2)}{4 E_{\nu}^2 m_N^2 x} F_{4,{\rm CC}}
%
- \frac{m_{\ell}^2}{E_{\nu} m_N} F_{5,{\rm CC}}
\Biggr)\, ,
\label{eq:nusig}
\end{eqnarray} 
for $x=Q^2/(2p\cdot q)$ and $y=p\cdot q/p\cdot k$, given $q=k-k'$ and $Q^2=-q^2$.
The sign of the $F_{3,{\rm CC}}$ term is $+(-)$ for neutrinos (antineutrinos). 

The CC structure function decomposition for $\nu p$ and $\bar\nu p$ scattering, at leading order in QCD for two generations of massless quarks, in terms of proton parton distribution functions labeled by their flavors are
\begin{eqnarray}
    F_{2,CC}&=& 2x (d + s +\bar{u}+\bar{c})\quad (\nu p)\\
    F_{3,CC}&=& 2(d + s - \bar{u}-\bar{c})\quad (\nu p)\\
    F_{2,CC}&=& 2x (u + c +\bar{d}+\bar{s})\quad (\bar\nu p)\\
    F_{3,CC}&=& 2(u + c - \bar{d}-\bar{s})\quad (\bar\nu p)   
\end{eqnarray}
where the structure functions and parton distribution functions depend on $x$ and $Q^2$. For isoscalar nucleon targets, we equate the up quark PDF in the neutron with the down quark PDF in the proton, and similarly for the up antiquark, down quark and down antiquark PDFs in the neutron. Next-to-leading order QCD corrections, quark mass corrections and target mass corrections modify the expressions, as summarized in e.g, ref. \cite{Kretzer:2002fr,Kretzer:2003iu}.

The limits of integration for eq. (\ref{eq:nusig}) are
\begin{eqnarray}
 && \frac{m_\ell^2}{2m_N(E_\nu -m_\ell)}\ \leq \  x\ \leq\ 1\ ,\\
&&a\ -\ b\ \leq \ y\ \leq \ a\ +\ b \ ,
\end{eqnarray}
where $a$ and $b$ are defined by
\begin{eqnarray*}
a & = & \Biggl[1-m_\ell^2\Biggl(\frac{1}{2m_NE_\nu x}+\frac{1}
{2E_\nu ^2}\Biggr)\Biggr]  
/(2+m_Nx/E_\nu)\ ,\\
b & =&  \Biggl[\Biggl(1-\frac{m_\ell^2}{2m_NE_\nu
  x}\Biggr)^2-\frac{m_\ell^2}{E_\nu ^2}\Biggr]^{1/2}  
/(2+m_Nx/E_\nu )\ .
\end{eqnarray*}

\section{Cross section tables}

Table \ref{Table:CS-Wnpi} -- \ref{Table:CS-W2p0} show the charged current cross sections of neutrino scattering with nucleons for muon neutrinos, tau neutrinos and their antineutrinos. 
We present the results evaluated using CKMT+PCAC-NT and the nCTEQ15 PDFs for $W_{\rm min} = m_N + n_\pi$, 1.4 and 2 GeV.

\begin{table*} [h]
\setlength{\tabcolsep}{10pt} 
\centering
\renewcommand{\arraystretch}{1.5}
\begin{tabular}{|c|c|c|c|c|}
\hline
\hline
\multirow{2}{*}{$E_\nu $ [GeV]} & 
\multicolumn{4}{c|}{$\sigma^{cc} (\nu N) /E_\nu~[10^{-38} {\rm cm^{2}/GeV}]$, \enskip $W_{\rm min}$ = $m_N + m_\pi$ } \\
\cline{2-5} 
& $\nu_\mu$ & $\bar{\nu}_\mu$ & $\nu_\tau$ & $\bar{\nu}_\tau$  \\
\hline
\hline
   1.000E+01 &    6.818E-01 &    3.006E-01&    1.335E-01 &    6.223E-02  \\ \hline
   1.259E+01 &    6.850E-01 &    3.062E-01&    1.800E-01 &    8.408E-02  \\ \hline
   1.585E+01 &    6.847E-01 &    3.108E-01&    2.282E-01 &    1.066E-01  \\ \hline
   1.995E+01 &    6.854E-01 &    3.152E-01&    2.758E-01 &    1.286E-01  \\ \hline
   2.512E+01 &    6.849E-01 &    3.185E-01&    3.215E-01 &    1.499E-01  \\ \hline
   3.162E+01 &    6.849E-01 &    3.218E-01&    3.639E-01 &    1.694E-01  \\ \hline
   3.981E+01 &    6.833E-01 &    3.237E-01&    4.028E-01 &    1.878E-01  \\ \hline
   5.012E+01 &    6.801E-01 &    3.253E-01&    4.369E-01 &    2.047E-01  \\ \hline
   6.310E+01 &    6.787E-01 &    3.276E-01&    4.682E-01 &    2.200E-01  \\ \hline
   7.943E+01 &    6.770E-01 &    3.298E-01&    4.953E-01 &    2.343E-01  \\ \hline
   1.000E+02 &    6.732E-01 &    3.305E-01&    5.176E-01 &    2.465E-01  \\ \hline
   1.259E+02 &    6.686E-01 &    3.309E-01&    5.359E-01 &    2.576E-01  \\ \hline
   1.585E+02 &    6.662E-01 &    3.327E-01&    5.524E-01 &    2.675E-01  \\ \hline
   1.995E+02 &    6.613E-01 &    3.331E-01&    5.652E-01 &    2.763E-01  \\ \hline
   2.512E+02 &    6.584E-01 &    3.340E-01&    5.759E-01 &    2.841E-01  \\ \hline
   3.162E+02 &    6.538E-01 &    3.345E-01&    5.834E-01 &    2.906E-01  \\ \hline
   3.981E+02 &    6.477E-01 &    3.338E-01&    5.881E-01 &    2.959E-01  \\ \hline
   5.012E+02 &    6.421E-01 &    3.347E-01&    5.920E-01 &    3.017E-01  \\ \hline
   6.310E+02 &    6.370E-01 &    3.348E-01&    5.938E-01 &    3.058E-01  \\ \hline
   7.943E+02 &    6.299E-01 &    3.342E-01&    5.933E-01 &    3.090E-01  \\ \hline
   1.000E+03 &    6.200E-01 &    3.329E-01&    5.892E-01 &    3.107E-01  \\ \hline
   1.259E+03 &    6.102E-01 &    3.321E-01&    5.843E-01 &    3.127E-01  \\ \hline
   1.585E+03 &    5.993E-01 &    3.296E-01&    5.771E-01 &    3.130E-01  \\ \hline
   1.995E+03 &    5.851E-01 &    3.260E-01&    5.669E-01 &    3.119E-01  \\ \hline
   2.512E+03 &    5.725E-01 &    3.237E-01&    5.568E-01 &    3.114E-01  \\ \hline
   3.162E+03 &    5.591E-01 &    3.216E-01&    5.447E-01 &    3.101E-01  \\ \hline
   3.981E+03 &    5.400E-01 &    3.157E-01&    5.282E-01 &    3.065E-01  \\ \hline
   5.012E+03 &    5.188E-01 &    3.090E-01&    5.092E-01 &    3.014E-01  \\ \hline
   6.310E+03 &    4.965E-01 &    3.021E-01&    4.887E-01 &    2.954E-01  \\ \hline
   7.943E+03 &    4.729E-01 &    2.939E-01&    4.664E-01 &    2.885E-01  \\ \hline
   1.000E+04 &    4.480E-01 &    2.857E-01&    4.424E-01 &    2.807E-01  \\ \hline
\hline
\end{tabular}
\caption{\label{Table:CS-Wnpi} The charged current cross sections of $\nu_\mu$, $\bar{\nu}_\mu$, $\nu_\tau$ and $\bar{\nu}_\tau$ scattering with isoscalar nucleon for $W_{\rm min}$ = $m_N + m_\pi$, evaluated using the nCTEQ15 PDFs and CKMT+PCAC-NT with $Q_0^2=4$ GeV$^2$.}
\end{table*} 


\begin{table*}[h]
\setlength{\tabcolsep}{10pt} 
\centering
\renewcommand{\arraystretch}{1.5}
\begin{tabular}{|c|c|c|c|c|}
\hline
\hline
\multirow{2}{*}{$E_\nu $ [GeV]} & 
\multicolumn{4}{c|}{$\sigma^{cc} (\nu N) /E_\nu~[10^{-38} {\rm cm^{2}/GeV}]$, \enskip $W_{\rm min}$ = 1.4 GeV }\\
\cline{2-5} 
& $\nu_\mu$ & $\bar{\nu}_\mu$ & $\nu_\tau$ & $\bar{\nu}_\tau$  \\
\hline
\hline
   1.000E+01 &    6.240E-01 &    2.495E-01&    1.033E-01 &    3.768E-02  \\ \hline
   1.259E+01 &    6.385E-01 &    2.641E-01&    1.522E-01 &    5.968E-02  \\ \hline
   1.585E+01 &    6.490E-01 &    2.768E-01&    2.037E-01 &    8.391E-02  \\ \hline
   1.995E+01 &    6.568E-01 &    2.875E-01&    2.544E-01 &    1.082E-01  \\ \hline
   2.512E+01 &    6.628E-01 &    2.959E-01&    3.036E-01 &    1.320E-01  \\ \hline
   3.162E+01 &    6.665E-01 &    3.038E-01&    3.486E-01 &    1.545E-01  \\ \hline
   3.981E+01 &    6.686E-01 &    3.092E-01&    3.905E-01 &    1.754E-01  \\ \hline
   5.012E+01 &    6.694E-01 &    3.140E-01&    4.271E-01 &    1.943E-01  \\ \hline
   6.310E+01 &    6.691E-01 &    3.183E-01&    4.596E-01 &    2.115E-01  \\ \hline
   7.943E+01 &    6.699E-01 &    3.223E-01&    4.885E-01 &    2.273E-01  \\ \hline
   1.000E+02 &    6.681E-01 &    3.247E-01&    5.124E-01 &    2.410E-01  \\ \hline
   1.259E+02 &    6.648E-01 &    3.265E-01&    5.320E-01 &    2.530E-01  \\ \hline
   1.585E+02 &    6.624E-01 &    3.290E-01&    5.489E-01 &    2.640E-01  \\ \hline
   1.995E+02 &    6.588E-01 &    3.302E-01&    5.626E-01 &    2.734E-01  \\ \hline
   2.512E+02 &    6.561E-01 &    3.318E-01&    5.737E-01 &    2.819E-01  \\ \hline
   3.162E+02 &    6.520E-01 &    3.328E-01&    5.818E-01 &    2.887E-01  \\ \hline
   3.981E+02 &    6.464E-01 &    3.324E-01&    5.868E-01 &    2.945E-01  \\ \hline
   5.012E+02 &    6.412E-01 &    3.333E-01&    5.911E-01 &    3.003E-01  \\ \hline
   6.310E+02 &    6.360E-01 &    3.339E-01&    5.929E-01 &    3.048E-01  \\ \hline
   7.943E+02 &    6.290E-01 &    3.335E-01&    5.924E-01 &    3.083E-01  \\ \hline
   1.000E+03 &    6.194E-01 &    3.325E-01&    5.887E-01 &    3.101E-01  \\ \hline
   1.259E+03 &    6.099E-01 &    3.317E-01&    5.839E-01 &    3.124E-01  \\ \hline
   1.585E+03 &    5.990E-01 &    3.293E-01&    5.768E-01 &    3.126E-01  \\ \hline
   1.995E+03 &    5.848E-01 &    3.256E-01&    5.666E-01 &    3.116E-01  \\ \hline
   2.512E+03 &    5.724E-01 &    3.235E-01&    5.567E-01 &    3.111E-01  \\ \hline
   3.162E+03 &    5.589E-01 &    3.214E-01&    5.445E-01 &    3.100E-01  \\ \hline
   3.981E+03 &    5.398E-01 &    3.155E-01&    5.281E-01 &    3.063E-01  \\ \hline
   5.012E+03 &    5.187E-01 &    3.092E-01&    5.091E-01 &    3.013E-01  \\ \hline
   6.310E+03 &    4.964E-01 &    3.019E-01&    4.886E-01 &    2.953E-01  \\ \hline
   7.943E+03 &    4.729E-01 &    2.939E-01&    4.663E-01 &    2.884E-01  \\ \hline
   1.000E+04 &    4.480E-01 &    2.857E-01&    4.424E-01 &    2.806E-01  \\ \hline
\hline
\end{tabular}
\caption{\label{Table:CS-W1p4}The charged current cross sections of $\nu_\mu$, $\bar{\nu}_\mu$, $\nu_\tau$ and $\bar{\nu}_\tau$ scattering with isoscalar nucleon for $W_{\rm min}$ = 1.4 GeV, evaluated using the nCTEQ15 PDFs and CKMT+PCAC-NT with $Q_0^2=4$ GeV$^2$.
}
\end{table*}


\begin{table*}[h]
\setlength{\tabcolsep}{10pt} 
\centering
\renewcommand{\arraystretch}{1.5}
\begin{tabular}{|c|c|c|c|c|}
\hline
\hline
\multirow{2}{*}{$E_\nu $ [GeV]} & 
\multicolumn{4}{c|}{$\sigma^{cc} (\nu N) /E_\nu~[10^{-38} {\rm cm^{2}/GeV}]$, \enskip $W_{\rm min}$ = 2.0 GeV} \\
\cline{2-5} 
& $\nu_\mu$ & $\bar{\nu}_\mu$ & $\nu_\tau$ & $\bar{\nu}_\tau$  \\
\hline
\hline
   1.000E+01 &    4.983E-01 &    1.586E-01&    4.849E-02 &    8.933E-03  \\ \hline
   1.259E+01 &    5.370E-01 &    1.860E-01&    9.893E-02 &    2.484E-02  \\ \hline
   1.585E+01 &    5.684E-01 &    2.104E-01&    1.545E-01 &    4.699E-02  \\ \hline
   1.995E+01 &    5.930E-01 &    2.321E-01&    2.104E-01 &    7.234E-02  \\ \hline
   2.512E+01 &    6.107E-01 &    2.493E-01&    2.652E-01 &    9.869E-02  \\ \hline
   3.162E+01 &    6.259E-01 &    2.661E-01&    3.161E-01 &    1.249E-01  \\ \hline
   3.981E+01 &    6.368E-01 &    2.787E-01&    3.636E-01 &    1.500E-01  \\ \hline
   5.012E+01 &    6.436E-01 &    2.891E-01&    4.048E-01 &    1.728E-01  \\ \hline
   6.310E+01 &    6.493E-01 &    2.985E-01&    4.415E-01 &    1.941E-01  \\ \hline
   7.943E+01 &    6.530E-01 &    3.057E-01&    4.730E-01 &    2.123E-01  \\ \hline
   1.000E+02 &    6.552E-01 &    3.119E-01&    5.004E-01 &    2.286E-01  \\ \hline
   1.259E+02 &    6.545E-01 &    3.158E-01&    5.223E-01 &    2.429E-01  \\ \hline
   1.585E+02 &    6.547E-01 &    3.207E-01&    5.412E-01 &    2.560E-01  \\ \hline
   1.995E+02 &    6.524E-01 &    3.237E-01&    5.565E-01 &    2.671E-01  \\ \hline
   2.512E+02 &    6.509E-01 &    3.266E-01&    5.685E-01 &    2.768E-01  \\ \hline
   3.162E+02 &    6.477E-01 &    3.282E-01&    5.777E-01 &    2.845E-01  \\ \hline
   3.981E+02 &    6.431E-01 &    3.292E-01&    5.835E-01 &    2.913E-01  \\ \hline
   5.012E+02 &    6.383E-01 &    3.307E-01&    5.882E-01 &    2.978E-01  \\ \hline
   6.310E+02 &    6.340E-01 &    3.318E-01&    5.908E-01 &    3.028E-01  \\ \hline
   7.943E+02 &    6.273E-01 &    3.317E-01&    5.908E-01 &    3.066E-01  \\ \hline
   1.000E+03 &    6.182E-01 &    3.309E-01&    5.875E-01 &    3.089E-01  \\ \hline
   1.259E+03 &    6.091E-01 &    3.305E-01&    5.831E-01 &    3.114E-01  \\ \hline
   1.585E+03 &    5.983E-01 &    3.283E-01&    5.761E-01 &    3.118E-01  \\ \hline
   1.995E+03 &    5.843E-01 &    3.249E-01&    5.660E-01 &    3.109E-01  \\ \hline
   2.512E+03 &    5.718E-01 &    3.230E-01&    5.561E-01 &    3.106E-01  \\ \hline
   3.162E+03 &    5.585E-01 &    3.209E-01&    5.441E-01 &    3.095E-01  \\ \hline
   3.981E+03 &    5.395E-01 &    3.151E-01&    5.277E-01 &    3.059E-01  \\ \hline
   5.012E+03 &    5.186E-01 &    3.090E-01&    5.089E-01 &    3.012E-01  \\ \hline
   6.310E+03 &    4.962E-01 &    3.017E-01&    4.884E-01 &    2.950E-01  \\ \hline
   7.943E+03 &    4.728E-01 &    2.937E-01&    4.662E-01 &    2.882E-01  \\ \hline
   1.000E+04 &    4.479E-01 &    2.856E-01&    4.423E-01 &    2.805E-01  \\ \hline
\hline
\end{tabular}
\caption{\label{Table:CS-W2p0}The charged current cross sections of $\nu_\mu$, $\bar{\nu}_\mu$, $\nu_\tau$ and $\bar{\nu}_\tau$ scattering with isoscalar nucleon for $W_{\rm min}$ = 2.0 GeV, evaluated using the nCTEQ15 PDFs and CKMT+PCAC-NT with $Q_0^2=4$ GeV$^2$.}
\end{table*} 

\bibliography{references}%

\end{document}